\documentclass[]{aastex631}
\pdfoutput=1 
\usepackage{amsmath,amstext,amssymb,mathtools,graphicx,color}

\usepackage{graphicx}	
\usepackage{amsmath}	
\usepackage{amssymb}	
\usepackage{xcolor}
\usepackage{lipsum}
\usepackage[normalem]{ulem}

\submitjournal{ApJ}

\shorttitle{Inhomogeneous Radiation Flow}
\shortauthors{Fryer et al.}

\graphicspath{{./}{figures/}}

\begin{document}

\title{Radiation-Hydrodynamics Effects in an Inhomogeneous Medium}

\correspondingauthor{Chris Fryer}
\email{fryer@lanl.gov}

\author[0000-0003-2624-0056]{Christopher~L. Fryer}
\affiliation{Center for Theoretical Astrophysics, Los Alamos National Laboratory, Los Alamos, NM, 87545, USA}

\author[0000-0001-5353-7483]{Paul A. Keiter}
\affiliation{Los Alamos National Laboratory, Los Alamos, NM, 87545, USA}

\author[0000-0002-4163-4830]{Vidushi Sharma}
\affiliation{Center for Theoretical Astrophysics, Los Alamos National Laboratory, Los Alamos, NM, 87545, USA}

\author[0000-0003-2393-3973]{Joshua Leveillee}
\affiliation{Center for Theoretical Astrophysics, Los Alamos National Laboratory, Los Alamos, NM, 87545, USA}

\author[0000-0002-0904-1250]{D.D. Meyerhofer}
\affiliation{Los Alamos National Laboratory, Los Alamos, NM, 87545, USA}

\author[0000-0002-4646-7517]{D. H. Barnak}
\affiliation{Laboratory for Laser Energetics, University of Rochester, NY, 14623, USA}

\author[0000-0003-3956-6506]{Tom Byvank}
\affiliation{Los Alamos National Laboratory, Los Alamos, NM, 87545, USA}

\author[0000-0002-7373-9303]{A. T. Elshafiey}
\affiliation{Los Alamos National Laboratory, Los Alamos, NM, 87545, USA}

\author[0000-0003-1087-2964]{Christopher J. Fontes}
\affiliation{Center for Theoretical Astrophysics, Los Alamos National Laboratory, Los Alamos, NM, 87545, USA}
\affiliation{Los Alamos National Laboratory, Los Alamos, NM, 87545, USA}

\author[0000-0001-7252-3343]{Heather M. Johns}
\affiliation{Los Alamos National Laboratory, Los Alamos, NM, 87545, USA}

\author[0000-0001-6849-3612]{P. M. Kozlowski}
\affiliation{Los Alamos National Laboratory, Los Alamos, NM, 87545, USA}

\author[0000-0002-7257-608X]{Todd Urbatsch}
\affiliation{Los Alamos National Laboratory, Los Alamos, NM, 87545, USA}



\begin{abstract}

Radiation flow through an inhomogeneous medium is critical in a wide range of physics and astronomy applications from transport across cloud layers on the earth to the propagation of supernova blast-waves producing UV and X-ray emission in supernovae.  Radiation interacts with matter driving hydrodynamic feedback that further alters the radiation characteristics (energy and angular distribution).  This paper reviews the current state of the art in the modeling of inhomogeneous radiation transport, subgrid models developed to capture this often-unresolved physics, and the experiments designed to improve our understanding of these models.  This paper focuses on simulations based on upcoming experiments designed to test this physics.  We present a series of detailed simulations (both single-clump and multi-clump conditions) probing the dependence on the physical properties of the radiation front (e.g. radiation energy) and material characteristics (specific heat, opacity, clump densities).  We find that, unless the radiation pressure is high, the clumps will heat and then expand, effectively cutting off the radiation flow.  The expanding winds can also produce shocks that generates high energy emission.  We compare our detailed simulations with some of the current subgrid prescriptions, identifying some of the limitations of these current models.  

\end{abstract}


\section{Introduction} 
\label{sec:intro}

For many physics and astrophysics applications, energy transport is often carried by photons, neutrinos and electrons.  In its generality, solving this energy transport problem requires modeling a seven dimensional problem where the spatial, angular and energy distribution of these carriers must be evolved.  Coupling this energy transport to matter drives further hydrodynamic instabilities, pushing the limits of our current modeling capabilities~\citep{2004rahy.book.....C}.  The difficulties in modeling radiation-hydrodynamics have led to many approximate methods that include simplifications in the radiation transport.  Approximations in the angular distribution include moment approaches (e.g. diffusion, variable Eddington factor and tensor methods).  The differences between these approaches lie in the order of the moments used.  More sophisticated, but typically more computationally expensive, approaches discretize the angle distribution using either spherical harmonics (e.g., $P_N$, \cite{1972NSE....49...10R,2000PhDT.......240B}) or discrete ordinate (e.g., $S_N$, \cite{1973erh..book.....P}) approaches.  Finally, Monte-Carlo approaches where particles/photons are represented by a set of packets have the ability to easily implement complex physics (e.g. aspherical scattering kernels, relativistic effects) but tend to be expensive and are hampered by noise from low packet numbers.  Because they have the potential to capture the full angular distribution, Monte Carlo and discretized methods are often termed ``higher-order" methods.

In astrophysics, most of the radiation-hydrodynamics calculations have focused on moment solutions, the simplest being the diffusion approximation~\citep{2004rahy.book.....C}.  Flux-limiters have been included to approximate the transition region between diffusive and free-streaming regimes~\citep{1981ApJ...248..321L,1982ApJS...50..115B}.  Radiation-hydrodynamics using flux-limited diffusion have been widely used in astrophysics, coupled to a broad range of hydrodynamics schemes from smooth particle hydrodynamics~\citep{1994ApJ...435..339H}, fixed grid and adaptive mesh refinement codes~\citep{2001ApJS..135...95T}.  Higher-order transport techniques coupled to hydrodynamics are gradually becoming more common.  For neutrino transport, astronomers are beginning to implement higher-order methods, developing spherical harmonic~\citep{2013JCoPh.242..648R}, discrete ordinate~\citep{1999A&A...344..533Y,2001PhRvD..63j3004L} and Monte-Carlo~\citep{2018PhRvD..98f3007F,2019ApJS..241...30M} approaches~\citep[for a review, see][]{2021ApJ...909..210A}.  Many of these methods couple to the hydrodynamics using operator-splitting techniques.  Although the supernova light-curve community uses a variety of higher-order transport techniques~\citep{2022A&A...668A.163B}, only a few efforts have used higher-order radiation transport methods coupled to matter~\citep[e.g.][]{2020ApJ...898..123F}.  

In many astrophysical applications, the implementation of the atomic physics can be extremely challenging, especially in systems where lines dominate the opacity.  Capturing line opacities, particularly in problems where there is considerable Doppler broadening can be challenging and a number of approaches ranging from pure Sobolev solutions\citep{1963trt..book.....S} to expansion opacities have been developed~\citep{1977ApJ...214..161K,1983ApJ...272..259F,1993ApJ...412..731E,1996AstL...22...79B,2003A&A...401...43W}.  Line-dominated opacities are extremely important for stellar winds, and this community has developed specialized opacity implementations and transport techniques to capture this physics~\citep{1980ApJ...241.1131C,2016MNRAS.458.2323K,2018A&A...611A..17S,2019A&A...631A.172D,2021A&A...648A..94L} although more-recent approaches have begun to introduce moment transport techniques such as flux-limited diffusion~\citep{2022A&A...665A..42M,2025A&A...696A.131N} and discretized angle solutions~\citep{2022ApJ...929..156G,2022ApJ...933..164G}.

These simplifying assumptions are all valid in specific conditions.  For example, in calculating supernova light-curves, the level of fidelity depends on how much detail we would like to learn from the observations.  The simplest assumption is that the radiation is trapped when the supernova blastwave is propagating through the star (pure hydrodynamics calculations are sufficient) and, after shock breakout, shocks are unimportant and pure transport calculations are sufficient.  For many supernova light-curve properties, these assumptions are good enough.  However, the most of the focus in the supernova community has been on the implementation of the opacities including out-of-equilibrium effects~\citep[see code comparison in][]{2022A&A...668A.163B}.  But it is becoming increasingly clear that shock interactions play a major role in supernova light-curves~\citep{2013ApJS..204...16F,2017ApJ...850..133D,2017hsn..book..403S,2019ApJ...876..148C}.  For the most part, these shock interactions are included using either recipes~\citep[e.g.][]{2010MNRAS.405.2113D,2017ApJ...850..133D,2025arXiv250115702N} or through radiation-hydrodynamics using the first moment (flux-limited diffusion) methods~\citep{1993A&A...273..106B,2013ApJS..204...16F}.  But a growing number of codes are adding improved transport schemes to hydrodynamic solutions, e.g. the variable Eddington Tensor code {\it Hydra}~\citep{2009AIPC.1171..161H}.

The promise of high-fidelity observations from upcoming ultraviolet missions, UltraSat~\citep{2024ApJ...964...74S}, UVEX~\citep{2021arXiv211115608K}, and potential X-ray missions~\citep{2024Univ...10..316A} mandates increasingly accurate calculations of shock interactions in supernova light-curves.  One of the primary effects seen in supernovae is the interaction with the radiation from the supernova with the clumpy circumstellar medium produced in stellar winds, common envelope mass ejection, and explosive stellar burning~\citep[for reviews, see][]{2017hsn..book..403S,2020ApJ...898..123F,2025arXiv250115702N}.  These interactions alter the observed supernovae in a number of ways.   For example, shocks at the boundary with clumps heats the ejecta, altering the temperature of the supernova.  the radiative heating can cause the denser clumps or shells to expand and this ``blow-off" that drives additional shock heating.  The higher temperatures achieved because of this shock heating is believed to produce the extended ultraviolet emission in supernovae~\citep{2017hsn..book..403S,2017ApJ...850..133D,2020ApJ...898..123F,2025arXiv250115702N,2022ApJ...931...15B}.  Understanding the role of this shock heating allows scientists to disentangle the heating from $^{56}$Ni decay from other sources (shock heating, long-lived central engines).  

The blow-off can also alter the observed spectra.  For example, the appearance of late-time H$\alpha$ lines in supernovae could be explained by interactions with common-envelope ejecta or a stellar companion~\citep[e.g.][]{2025arXiv250408889R}.  Depending on the composition and distribution of the clumpy-material, this blow-off could alter the propagation of the radiation front.

This paper investigates the radiation flow through these clumpy media, studying the physics effects on the radiation transport and coupling between radiation and hydrodynamics.  These studies will focus on leveraging laboratory experiments to gain insight into the physics.  These experiments can provide insight into the nature of blow-off and its effect on the opacity in a radiation flow.  Although this is an important part of the physics in astrophysical applications such as supernovae, it does not study all of the needed physics.  For example, while velocity broadening and line opacities in a velocity gradient are important in supernovae, pressure broadening is far more important in these experiments.  The implementation of the opacities is more straightforward in the experiments.  In our study, we focus on the coupling between the radiation and the hydrodynamics and the effects of the blow-off. 

In addition to the aforementioned numerical approximations in the transport, radiation-hydrodynamics solutions are typically under-resolved.  Capturing the effects of heating and blow-off in a solution requires resolving the clumps.  Numerical diffusion in hydrodynamics can both artificially transport material and introduce a viscosity to problems in fluid flow~\citep{2004rahy.book.....C}.  Improved discretization techniques (e.g. techniques that include subgrid information) and higher resolution can minimize the effects of this false diffusion.  Transport can also experience an effective numerical diffusion, a.k.a. teleportation, where radiation artificially diffuse through a material.  As with hydrodynamic diffusion, these effects can be mitigated with subgrid models, e.g.``tilt'' functions~\citep{2020JCoPh.41209405P}.

These resolution effects become more difficult when a radiation flow passes through an inhomogeneous medium where the inhomogeneities or clumps can not be fully resolved.  A number of subgrid models have been developed to try to capture the physics of radiation flow through an inhomogeneous medium.  But, as we shall see in this paper, these recipes capture only one aspect (the increase in opacity) of the inhomogeneous radiation flow under the assumption that no hydrodynamic feedback occurs.  This paper studies the physical effects in of radiation flow through an inhomogeneous medium including the increase in opacity, but also on effects that alter the angular and spectral features of the radiation such as shock heating as the blow-off from the clumps interact with their surroundings.  For supernova observations, this blow-off and the shocks may be the most important effect of inhomogeneous radiation flow.  We also study the stochasticity of the interactions.  These studies are designed to indentify the that must be captured in a subgrid model to fully incorporate the effects of radiation flow in an inhomogeneous medium. 

Although we focus on photon radiation-transport in this paper, manny of the conclusions also apply to other energy carriers including neutrinos and electrons scaled by the transport regime (mean free path) and emission/deposition properties.  Section~\ref{sec:app} reviews a set of applications where radiative transport through an inhomogeneous medium is important, discussing which physical effects are most critical in each application.  Section~\ref{sec:soa} reviews the current state of both modeling and experiment to study inhomogeneous radiation flow.  In this paper, we focus on detailed radiation-hydrodynamics calculations to better understand the physics behind radiation flow and Section~\ref{sec:simulation} describes our initial conditions and the computational methods used in this paper.  Section~\ref{sec:single} focuses on the basic physics behind radiation flow across a single clump.  Section~\ref{sec:multi} studies the more complex interactions in a multi-clump medium.  We conclude with a discussion of how the physics studied in these simulations affects the different applications.

\section{Applications}
\label{sec:app}

Radiation transport is critical in a wide range of applications including oncological treatments, radiation shielding, nuclear reactors, astrophysics, and planetary atmospheres; for a review, see~\citep{byvank23}.   In some cases (particularly in astrophysics), the deposition of energy and momentum from this radiation plays a critical role in the evolution of the phenomena.  For example, momentum deposition from radiation is key in understanding radiation-driven outflow from massive stars.  Astronomers are now realizing that the smooth, line-driven winds~\citep{2008A&ARv..16..209P} are too simple to explain all of the features in the stellar wind profile.  Explosive shell burning, opacity-driven instabilities and pressure waves can all produce bursts of mass ejection in a star, producing inhomogeneities in the circumstellar medium including both shells and clumpy media~\citep{2006ApJ...647.1269F,2014ApJ...792L...3H,2016MNRAS.458.1214Q}.  Studies of line-driven winds indicate that even these, relatively quiescent, outflows, can produce large inhomogeneities in the circumstellar medium~\citep{1984ApJ...284..337O,2008A&ARv..16..209P,2018A&A...611A..17S,2018Natur.561..498J,2019MNRAS.485..988O}.  In particular, clumping in the line-driven wind can alter the radiation and mass outflow from these winds~\citep{2018A&A...611A..17S,2018MNRAS.475..814O}.  Understanding the interaction between radiation and these clumps determines both this mass loss and the evolution of these clumps.  Especially in line-driven winds, the implementation of the opacity, can play a critical role in determining the effects of these flows.  This paper will focus on transport effects and not on opacity implementations.  \cite{2004rahy.book.....C} reviews these methods in detail.

The clumpy structure produced in stellar mass ejection provides the initial conditions through which supernova explosions propagate.  Radiation is typically trapped in the material shock until the supernova breaks out of the star.  It is believed that the edge of the star is marked by a sharp density break as the stellar profile transitions to a stellar wind profile.  In simple spherically symmetric models, the radiation in the shock rapidly transitions from being trapped in the material flow to freely streaming out of the shock~\citep{2017hsn..book..967W}.  This effect depends sensitively on the structure of the edge of the star as well as asymmetries in the supernova blastwave (from the explosion and the stellar envelope) which complicate this picture~\citep{2015ApJ...805...98B,2017ApJ...845..103L,2020ApJ...898..123F,2021MNRAS.508.5766I}.  Especially in Wolf-Rayet stars where shock breakout actually occurs in the circumstellar medium, not the edge of the star, shock breakout can be drastically modified by the radiation flow through the clumpy wind material~\citep{2020ApJ...898..123F}.  Interactions between the radiation and the clumps disrupt the clumps and the ablation of these clumps produce shocks that drive high-energy emission~\citep{2020ApJ...898..123F}.

Beyond shock breakout, interactions of the radiation flow with the circumstellar medium can dramatically alter the observed supernova light-curves.  Interactions of the radiation with the circumstellar medium drive many of the features seen in type II narrow-line supernovae.  Most of the current studies focus on the interaction with a spherically symmetric shell~\citep[e.g.][]{2012ApJ...751...92R}, but it is likely that the circumstellar medium is much more heterogeneous.

Transport of radiation through a dusty medium is another application where including the inhomogeneities of the clumps (dust particles) can alter the propagation of radiation and albedo of a dust cloud~\citep{1990A&A...228..483B,1996ApJ...463..681W,2000ApJ...528..799W,2003MNRAS.342..453H}.  This phenomenon plays a role in analyzing observational data of these dust clouds, the penetration of energy in molecular clouds, and the destruction of dust.  These effects, in turn, can alter star formation in molecular clouds.  The propagation of radiation around dense clumps in molecular clouds can cause the clumps to implode, driving star formation~\citep{1980SSRv...27..275K,1982ApJ...260..183S,2011ApJ...736..142B,2014ApJ...780...51P}.  This inhomogeneous transport should also affect the interpretation of nova and supernova observations, but less work has been done in this field.

Climate and environmental studies have also studied inhomogeneous radiation flow.  Understanding the transport through vegetation canopies is important both for agriculture and climate modeling~\citep{2003JQSRT..77..373K,2007JQSRT.107..236S,GANAPOL1999153}.  Climate scientists have studied radiative transfer through inhomogeneous cloud cover~\citep{2004AtmRe..72..223C,2016JGRD..121.8583H} and recipes to mimic that transfer are used in many climate codes~\citep{2013AGUFM.A43B0256W,2016JGRD..121.8583H}.  With improving satellite data, tests of these models become increasingly stringent, driving advances in this transport.   

Studies of transport through an imhomogeneous or stochastic medium is not limited to photons.  A growing number of charged particle transport studies, particularly of cosmic rays (high energy electrons, protons and ions), have focused on the radiation flows across single clumps~\citep{2019MNRAS.489..205W,2021ApJ...913..106B} and turbulent media~\citep{2022ApJ...931..140H}.  These studies focused both on the effect of the clumpy media on the flow of radiation and the effect of the radiation on the clumps, including the energy and momentum deposition of the radiation into the clumpy media.

\section{Current Understanding}
\label{sec:soa}

\subsection{Inhomogeneous Transport Solutions}
\label{sec:transportsol}

The effects of radiation transport through inhomogeneous media depends on the properties of the inhomogeneities.  Studies of inhomogeneities generally consider a 2-component medium consisting of an optically-thin radiation flow region with optically-thick embedded clumps.  \footnote{Although not strictly appropriate for radition-driven winds~\citep{2022A&A...665A..42M,2024A&A...684A.177D}, the embedded clumps do work for models where clumpy hydrogen material from a common envelope mass-ejection phase is surrounded by a subsequent radiatively-driven wind phase.}  Developing a subgrid model to capture the physics of this radiation flow depends upon the exact physics that must be captured for a given application.  Here we discuss three different physical effects:
\begin{itemize}
    \item {\it Radiation Propagation:}  The clumps will alter the propagation of the radiation front.  If the optically-thin region is in the free-streaming regime, the optically-thick clumps will reduce the radiation that propagates through this region.  If the optically-thin region is close to the transport regime, the clumps can cause the radiation front to decelerate.
    \item {\it Radiation Direction:}  The clumps not only affect the propagation, but also the angular distribution of the radiation as the clumps heat and re-emit.  This effect is important for applications but only noticed in codes that capture this physics, e.g. variable Eddington tensor, spherical harmonics ($P_N$), discrete ordinate ($S_N$), or Monte Carlo schemes.
    \item{\it Hydrodynamic Feedback and Shocks:}  When the clumps are heated, they will expand.  In supernovae, the shocks produced by these expanding shocks are critical in understanding early-time ultraviolet and X-ray emission including shock breakout as well as the peak optical emission~\citep{2016ApJ...820...23G,2017hsn..book..403S,2017ApJ...850..133D,2020ApJ...898..123F}.
\end{itemize}  
Ideally, any subgrid model would capture all of this physics.  But, as we shall see, most models have focused on radiation propagation and, to a lesser extent, angular distribution of the radiation.  By not including the hydrodynamic feedback, these methods can underestimate the opacity, preventing an accurate model for the radiation propagation (see Section~\ref{sec:multi}). To our knowledge, no subgrid model includes the radiation-induced shocks and the subsequent shock heating.

Here we review a couple of the current subgrid models.  Throughout this paper, we will refer to the ``flow" region as the optically-thin region through which the radiation front propagates, and the clump region as the denser, optically-thick clumps around which the radiation flows.  In experiments, the flow region is often composed of a silicate foam and the clumps are inclusions in that foam.  In many astrophysical applications, the flow region is simply the lower density region where the clumps are produced by dense turbulent eddies.  The effect of the clumps depends on their densities, opacities, specific heats, geometry, sizes, and covering fractions.  It depends on the transport regime:  diffusion limit, transport regime, free-streaming limit of the radiation flow.  In the free-streaming limit, dense clumps can completely alter the nature of the flow.  For instance, inhomogeneities in atmospheric clouds can lead to an increase in the albedo of the cloud layer, decreasing the amount of radiation that reaches a planet's surface.  In the diffusive regime, radiation is more able to flow around clumps.  Transport-regime conditions lie in between these two extremes.

Many of the solutions developed for radiation transport in inhomoegeneous media are most appropriate in the diffusion limit.  These methods often use Markov-process solutions~\citep{1986JQSRT..36..557V,1989TTSP...18..287M,1994JQSRT..51..467S,1995JQSRT..54..779S,2003JQSRT..81..451S,2003JQSRT..77..373K,2006JQSRT.101..269O,2007JQSRT.104...86O,2014arXiv1410.8200D} with Poisson distributions of the inhomogeneities.  Some studies derived solutions assuming specific properties of the inhomogeneities~\citep{1988JQSRT..39..333V}.  Other studies focused on understanding the solution limits~\citep{1989TTSP...18..287M, 2001JQSRT..70..115M}.  The simplifications required in these analytic or semi-analytic solutions drove scientists to develop increasingly sophisticated modeling tools to capture radiation-flow in heterogeneous media~\citep{2011JQSRT.112..599B,osti_1824766,2021EPJWC.24704009L}.  Many of the studies focused on developing reduced-order models that rely on a subset of the relevant physics (e.g. mean free path with respect to clump size, number density of clumps, clump spacing, etc.).

A straightforward approach to introduce inhomogeneites in reduced order techniques is to derive an effective opacity in the region with inhomogeneities.  For example, the approach of \cite{2005JQSRT..90..131P,2007JQSRT.104...86O} included an additional effective scattering opacity as well as a modification to the absorption opacity for an inhomogeneous medium for a purely absorptive medium  (i.e. the scattering opacity is much lower than the absoprtion opacity).  The effective absorption opacity ($\sigma_{\rm a, eff}$) is given by:
\begin{equation}
\sigma_{\rm a, eff} = 1/(p_1/\sigma_{a1} + p_2/\sigma_{a2})
\label{eq:sigabs}
\end{equation}
where $p_{1,2}$ are the covering fractions or covering fractions of the clump and surrounding medium respectively and $\sigma_{a1,a2}$ are the equivalent absorption opacities for these materials.  They included an effective scattering opacity ($\sigma_{\rm s, eff}$):
\begin{equation}
\sigma_{\rm s, eff} = \nu^2/[\tilde{\sigma} (1 + \lambda_c \tilde{\sigma})]
\label{eq:sigscat}
\end{equation}
where $\tilde{\sigma}= p_1 \sigma_{a1} + p_2 \sigma_{a2}$ is the average opacity weighted by the covering fraction, $\nu^2=p_1 p_2 (\sigma_{a1}-\sigma_{a2})^2$ is the unconditional variance in the opacity, and $\lambda_c = \lambda_1 \lambda_2/(\lambda_1+\lambda_2)$ is the geometric mean of the chord lengths, where the chord length is defined by the length of a chord covered by each material.

\begin{figure}[ht]
\includegraphics[width=4.1in]{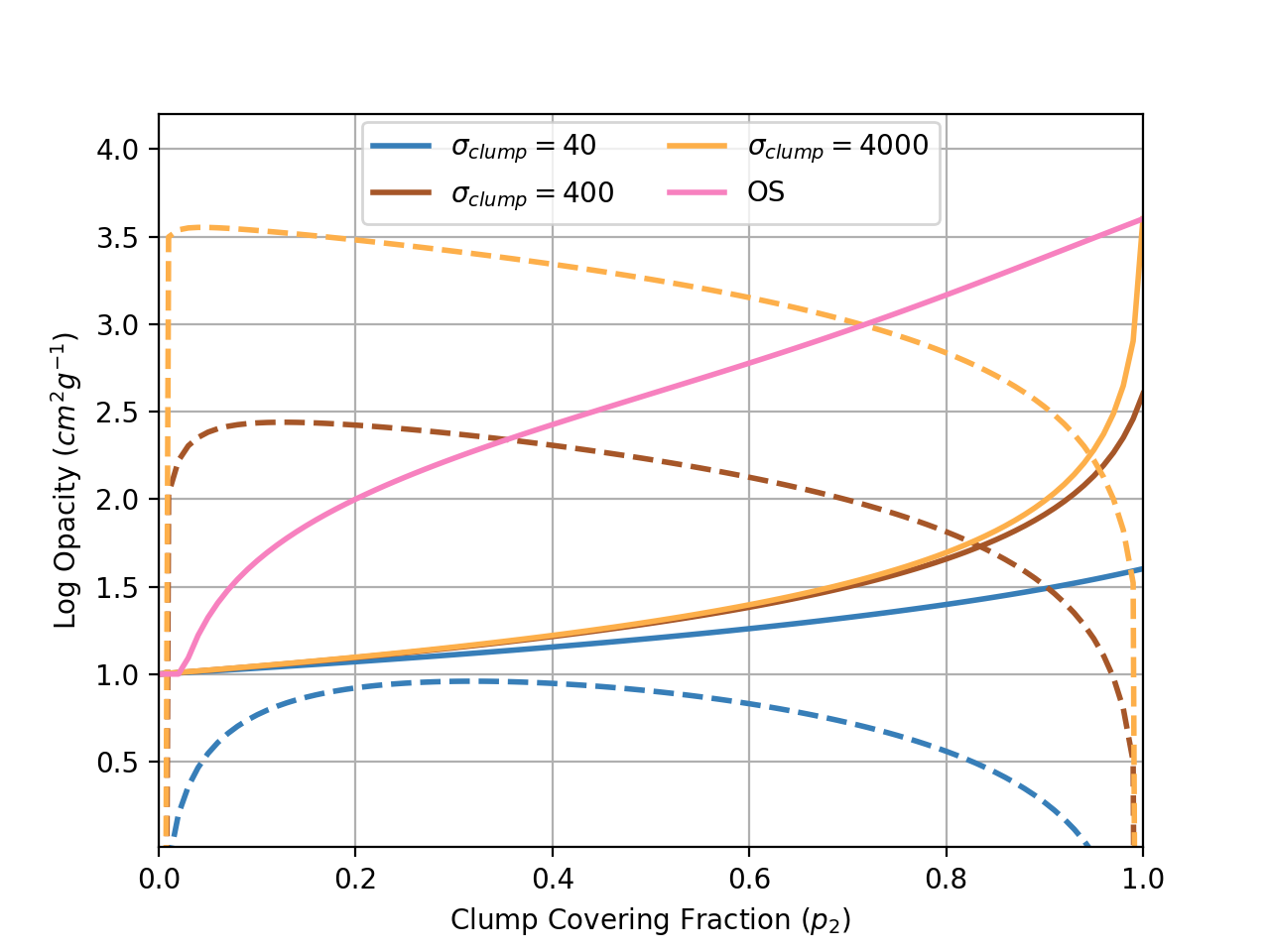}
    \caption{Effective absorption (solid) and scattering (dashed) opacities for three values of clump opacity (in ${\rm cm}^2 \, {\rm g}^{-1}$) as a function of clump covering fraction.  The opacity of the flow region is $10 {\rm \, cm^2 \, g^{-1}}$ and the density of the flow, clump regions are $0.05 {\rm \, g \, cm^{-3}}, 5  {\rm \, g \, cm^{-3}}$ respectively.  The effective absorption opacity does not increase significantly until the covering fraction exceeds 0.8.  The ``OS'' model corresponds to the ~\cite{2018MNRAS.475..814O} formula (equation~\ref{eq:sigowo}).
    However, the effective scattering opacity can be high if the clump opacity is high compared to the opacity in the flow region.}
    \label{fig:opac}
\end{figure}

Using this prescription, we can calculate the effective absorption and scattering opacities as a function of the covering fraction of the clumps.  Figure~\ref{fig:opac} shows the effective absorption and scattering opacities under this formalism as a function of the clump covering fraction.  Although the effective absorption cross section does not increase until the clump covering fraction is large, the effective scattering term quickly approaches the value of the clump opacity.  This scattering term can drastically lower the propagation speed of the radiation.  It also provides a means to capture the angular redistribution from the clumps.

For stellar winds, \cite{2018MNRAS.475..814O} developed a mechanism that is a direct modification to the total opacity based on the porosity of the clumps.  In this model, the effective opacity ($\sigma_{\rm eff}$) is given by:
\begin{equation}
    \frac{\sigma_{\rm eff}}{\sigma_{\rm clump}} =  \frac{1-e^{-\tau_h}}{\tau_h}
    \label{eq:sigowo}
\end{equation}
where $\tau_h = \langle \rho \sigma \rangle h$ where $h = (f_{\rm clump}-1) l_{\rm clump}$, $f_{\rm clump} = (l_{\rm sep}/l_{\rm clump})^3$, $l_{\rm clump}$ is the characteristic size of the clump and $l_{\rm sep}$ is the average separation of the clump.  $f_{\rm clump}$ is equivalent to the covering factor of the clumps ($p_2$) used in equation~\ref{eq:sigabs}.  As the covering factor approaches 1, both models set the opacity to the clump opacity.  In \cite{2018MNRAS.475..814O}, the opacity of the flow region is assumed to be zero and, as the filling or covering factor approaches 0, the opacity converges to zero.  To compare to the \cite{2007JQSRT.104...86O} solution, we set the final opacity to $\sigma_{\rm flow} + \sigma_{\rm eff}$ and include this opacity in Figure~\ref{fig:opac}.  This method focuses on capturing a mean opacity, but without an effective scattering term, is not able to capture the angular redistribution.

Neither of the above techniques include the shock heating and the effect it has on the radiation spectrum.  More work is needed to compare the full physical effects of clumps on radiation flows.

Most of these past studies focused on transport-only solutions or solutions where radiation/hydrodynamic effects can be minimized.  For example, \cite{2005JQSRT..90..131P} tested their scheme using a radiative transfer code.  Although the transfer could heat the material, these models were not coupled to a hydrodynamics package to model the subsequent expansion from the pressure gradient produced by the heating.  As such, the solutions typically only consider the relative opacities of the different clumps.  More detailed studies included the effects of the size scale of the clumps (with respect to the photon mean free path).  This effect becomes increasingly important as we move from the diffusive to the free-streaming transport regimes.

\subsection{Experiments and Scaling}
\label{sec:soa-exp}

A number of experiments studying radiation transport in inhomogeneous media already exist.  These experiments are primarily focused on studying the propagation of radiation through this media.  But a number of diagnostics have been improved to get at the effect of the inhomogeneities on the energy and angular distributions of the the propagating radiation.  Dante X-ray spectrometers are often placed at the edge of the target to measure the coarse spectrum of the emerging radiation front~\citep{2006RScI...77jE518S,2020PhPl...27k2709B}.  Absorption spectroscopy is being used to measure the atomic states, and hence temperature, of the material as the radiation front moves through the inhomogeneous medium~\cite{2023RScI...94b3502J}.  These measurements allow experimentalists to more directly observe the response of the dense material to being heated by the radiation. These outflows drive shocks that can then be probed by the spectral measurements.  Bear in mind, however, that these experiments can not be used to test opacity implementations in high-velocity media (e.g. expansion opacity recipes).

The scaling of experiments to applications has been discussed extensively in the literature~\citep{2005Ap&SS.298...49D,2011ApJ...730...96F,2017HEDP...22...37K,2020HEDP...3500738F,2024PhPl...31c2304K}.  Radiation transport is well understood in extreme limits:  in the optically-thick diffusion regime and the optically-thin free-streaming regime.  What many radiation experiments probe is that intermediate transport regime.  The uncertainties in numerical implementations of the Boltzmann transport equation are greatest in this regime.  The primary experimental scaling parameter tying laboratory experiments to applications in this transport regime is the mean free path: e.g. number of mean free path across the target, number of mean free paths across a clump, number of mean free paths separating clumps.  By using either light foam or gas materials in the flow region, experimentalists can vary these properties.

However, the goal is instead to study the reaction of the material to radiation.  These experiments seek to test the hydrodynamics and transport schemes as well as the method used to couple these two physics models.  The amount of outflow of a radiatively-heated dense clump/shell depends upon the characteristics of the clumps (e.g. densities, specific heats).  But if we are more interested in testing numerical artifacts in numerics of radiation-hydrodynamics coupling:  e.g. the outflow will depend on differences in the treatment between Lagrangian and Eulerian schemes~\citep[e.g.][]{2018ACP....1816619K} and on effects of teleportation (or related numerical diffusion) in transport~\citep{2010PhDT........16C}.  With such studies direct scaling is not critical.  Indeed, ideally the experiment is designed to exacerbate these effects so that we can more easily identify the needed improvements to the algorithms.

With these scaling issues in mind, we review a set of existing experiments.  For code validation, it is important to understand the details of the experiments to both understand the uncertainties in the initial conditions, uncertainties in the physics assumptions and limitations of the diagnostics.  The following detailed discussions give examples of the difficulties in producing accurate validation tests.


\cite{shaoen2005supersonic} performed experiments investigating supersonic radiation propagation in a low-density heterogeneous medium.  These experiments investigated radiation propagation in a low-density plastic foam target as well as in a Cu-doped foam.  The Cu was added by means of small Cu particles and achieved roughly a 2.14\% atomic fraction.  The actual placement of the particles is difficult to control.  The foams were viewed face-on with a trichromatic streaked x-ray spectrometer (TCS).  The TCS consisted of a three-imaging-pinholes array and a three-transmission-grating array coupled with an x-ray streak camera.  The TCS observed the emission from two different photon energies, 210~eV and 840~eV. A soft x-ray spectrometer observed the soft x-ray spectrum of the hohlraum, from which a temperature could be inferred. Finally, a transmission grating spectrometer was used to measure the time-integrated spectrum from either the hohlraum or the foam sample. A delay in the radiation breakout was observed when comparing the Cu-doped foam with the pure foam case for emission at 210 eV; however, this trend was reversed when comparing emission at 840 eV. The opacity for both the pure foam and the Cu-doped foam was estimated at each of these energies. 
This study concluded that differences in the opacities caused the different measured radiation flow speeds.  But as we shall see here, the flow may also depend on the blow-off of the inclusions in the flow region.


Subsequent experiments by \textcite{2008PhPl...15e6901K} examined radiation with multiple particle sizes. These experiments used a gold hohlraum as a radiation source to drive a supersonic heat wave into a foam package. The inhomogeneous medium was a low-density plastic foam  with Au particles. The Au fraction from the particles ranged from 0 to 14\% by atomic fraction. This type of characterization allows one to visualize the distribution of Au particles throughout the foam and to determine if the particles are clumping together, thereby resulting in a different distribution than originally thought.  As we shall see in Section~\ref{sec:multi}, characterization of these foams is absolutely crucial for understanding the radiation transport. A detailed description of the complete characterization of these particular foams is found in \textcite{2008PhPl...15e6901K}.

Three different regimes were studied in these experiments, corresponding to no particles (or the homogeneous situation) and particles that are sub-micron and 6~microns in diameter. The position of the radiation front was measured by observing the self-emission of the foam with a soft x-ray imager \cite{ze1992new}. The soft x-ray imager observed a narrow bandwidth of radiation, peaking around 270~eV. The position of the radiation front was measured in each experimental case and compared with simulations. It should be noted that while foam packages were observed face-on in many previous radiation flow experiments, these measurements were made side-on in order to allow multiple measurements to be made and a time history to be compiled.

In order to demonstrate the ability to correctly simulate and understand the radiation propagation with this particular foam, measurements were first obtained with a homogeneous foam.
Next, measurements of the position of the radiation front were made in inhomogeneous foams and then compared with two types of simulations: simulations of the pure foam case and simulations using an atomic-mix model. Although the atomic-mix model matches the smallest particles well, neither model could explain the radiation front position for the 6~micron diameter particles.  The Pomraning model \citep{pomraning1998renormalized}, a precursor to the model described in Section~\ref{sec:transportsol}, was then applied to the simulations. This model assumed a mean particle size of 6~microns in diameter and the particles were in pressure equilibrium with the foam. The mean opacity was determined based on the material properties at each timestep in the calculation. The mean opacity was then divided by the opacity of the homogeneous foam to determine a correction factor to apply. This correction factor depended on the temperature of the Au-foam mixture and therefore changed in time as the temperature of the system evolved. Within the uncertainties of these experiments, the Pomraning model provided a good fit to the data.

These experiments have demonstrated the potential of laboratory experiments to probe the different physical aspects of the inhomogeneous radiation flow problem.  As increasingly sophisticated diagnostics are developed, these experiments will provide a means to test radiation flow and its hydrodynamic feedback.

\section{Physics of Inhomogeneous Flows}
\label{sec:soa-physics}

In this section, we present a wide set of calculations of radiation flow across an inhomogeneous medium, using simplified conditions to better understand the physics.  Section~\ref{sec:simulation} discusses the basic models used followed by two sections focusing on single- (Section~\ref{sec:single}) and multi-clump (Section~\ref{sec:multi}) physics effects.

\subsection{Simulation Tools and Initial Conditions}
\label{sec:simulation}

For our calculations, we use conditions close to that of the XFOL and XFLOWS experimental campaigns led by LANL~\citep{2023RScI...94b3502J, byvank23} and the {\it Cassio} code developed under LANL's Advanced Simulation and Computing program ~\citep[for details, see][]{2008CS&D....1a5005G,urbatsch2006milagro,2020ApJ...898..123F}.  The {\it Cassio} code models hydrodynamics using a cell-based adaptive mesh refinement scheme with a two-shock approximate Riemann solver using the methods described in the xRAGE code~\citep{2008CS&D....1a5005G}. The code has been verified against a variety of analytic test problems, the most relevant for this problem being the Sedov blast wave~\citep{2008CS&D....1a5005G}.  {\it Cassio} has several implementations of multi-material physics.  For our calculations, we use the default prescription.  This prescription assumes that when multiple materials are in a zone, the materials are in pressure-temperature equilibrium.  The relative density of the materials is determined by the specific heat capacities of the materials.  In this prescription, even if the mass fractions are the same, the materials can have different volume fractions and different densities.  The inferred densities are used in calculating the opacities.  Multi-material solutions such as this are not commonly used in astrophysics (for monatomic gases, the specific heat capacities are all 3/2), but are critical in a lot of applied physics, experiments and engineering solutions~\citep{2025NSE...199S.941H}.  Time-dependent radiation transport can be modeled using a variety of transport solutions including flux-limited diffusion~\citep{2008CS&D....1a5005G}, discrete ordinate ($S_N$)~\citep{2020ApJ...898..123F}, and Implicit Monte Carlo transport~\citep{urbatsch2006milagro} using a multi-group energy resolution.  

The {\it Cassio} code has been used extensively in the laboratory experimental community for a wide variety of problems studying hydrodynamics or radiation-hydrodynamics~\citep[e.g.][]{2016RScI...87kE337J,2017PPCF...59a4050F,2018PhRvL.120b5002F,2020PhRvL.124o9901F,2020HEDP...3500738F,2020PPCF...62g4001F,2021HEDP...3900939J,2022PhPl...29h3302C,2023RScI...94b3502J,2023HEDP...4601023F}.  In addition to many of these validation tests, {\it Cassio} has been verified against other experimental codes~\citep[e.g.][]{2013HEDP....9...63F,2014PhRvE..90c3107F}. {\it Cassio} and its related flux-limited diffusion version xRAGE have been applied to a number of radiation-hydrodynamics problems in astrophysics, mostly studying supernova light curves~\citep{2009ApJ...707..193F,2010AIPC.1294...70F,2010ApJ...725..296F,2013ApJ...762L...6W,2013ApJ...768...95W,2013ApJ...768..195W,2013ApJS..204...16F,2013ApJ...773L...7F,2013ApJ...774...64W,2013ApJ...777...99W,2013ApJ...777..110W,2013ApJ...778...17W,2014ApJ...781..106W,2014ApJ...797....9W,2014ApJ...797...97S,2015ApJ...805...44S,2020ApJ...898..123F} and has been verified against other astrophysical codes as well~\citep{2014JCoPh.275..154J}.  

Our calculations used a binned opacity approach similar to that used in many supernova codes including {\it SuperNu}~\citep{2013ApJS..209...36W} and {\it STELLA}~\citep{1992SvAL...18...43B}.  The binned approach has been compared to other line sampling and expansion opacity methods~\citep{2014ApJS..214...28W,2020MNRAS.493.4143F,2022A&A...668A.163B} with reasonable agreement.  The binned opacities are obtained from the OPLIB database~\citep{colgan_oplib} calculated with the Los Alamos suite of atomic physics codes~\citep{LANL_suite}.  The bins are created using a Rosseland mean approach, but a Planck averaging can be used (and is sometimes implemented) in the emission terms of the transport equation.  For most of our calculations, we use {\it Cassio}'s discrete-ordinate, $S_N$, transport scheme with an $N=8$ quadrature set and 71 energy groups\footnote{For more details on the $S_N$ method implemented in {\it Cassio}, see~\cite{carlson1965transport,2020ApJ...898..123F}}.  The transport solution is placed on the same grid as the hydrodynamics calculations.  The coarse grid resolution is $0.6 \, {\rm \mu m}$ and we use only 2 levels of refinement.  We discuss comparisons to other transport schemes and grid resolutions below.

\begin{figure}[ht]
\includegraphics[width=5.1in]{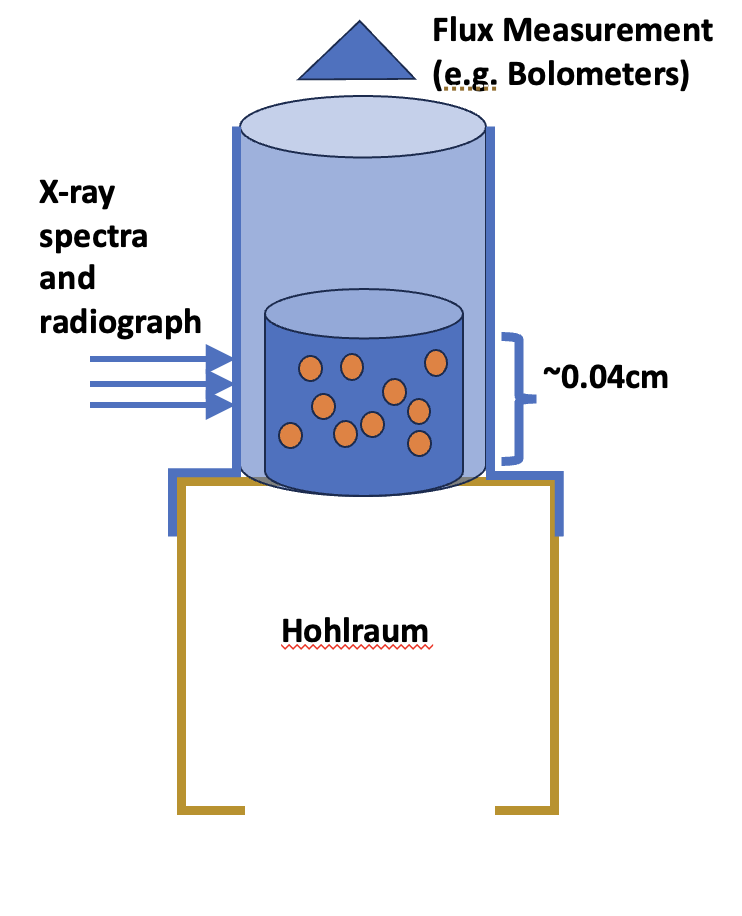}
    \caption{Simplified diagram of the experiment XFOL experiment~\citep{2023RScI...94b3502J} designed to measure radiation flow through an inhomogeneous medium.  Lasers heat the gold-line hohlraum cavity, producing a hot plasma that drives a radiation front through a target.  Dense inclusions are added into a low-density foam target that is roughly 0.04\,cm in size.  The radiation front can be probed both using a radiograph to measure the position of the front or a spectrograph to constrain the temperature and composition.  The escape of the radiation front from the top of the target can be used to measure the radiation flux versus time (the current version of this experiment does not have a detector for this measurement).}
    \label{fig:exp}
\end{figure}

The initial conditions for our simulations are guided by active XFOL and XFLOWS experimental campaigns~\citep{2023RScI...94b3502J, byvank23} that builds upon past Pleiades and COAX radiation flow studies~\citep{2020HEDP...3500738F,2021HEDP...3900939J,2022PhPl...29h3302C}.  In these experiments, a laser-driven hohlraum (either at the Omega laser facility in Rochester~\citep{1995RScI...66..508B} or the National Ignition Facility at Lawrence Livermore National Laboratory~\citep{2007ApOpt..46.3276H} produces a radiation front that then propagates through a target. This experiment consists of a target placed on top of a hohlraum (Figure~\ref{fig:exp}).  Lasers fired into the hohlraum create a hot, radiation-dominated plasma that then propagates through the experimental target.  The power in this drive peaks quickly and then decays slowly over the course of our few-ns simulation.  In our multi-clump, high-drive models, we also consider a simplified flat power (or temperature) for this drive.  The target is a silicon dioxide foam filled with vanadium-oxide inclusions.  In this paper, we focus solely on modeling the target, using a radiation source term at its base to represent the hohlraum-produced radiation flow.  The radiation source is a fit to a particular set of hohlraum models with a radiation temperature that is relatively flat over the course of the simulation (peaking at 150eV at 1\,ns).  The target is modeled in cylindrical symmetry with a radius of 0.04\,cm and a height of 0.08\,cm.

\subsection{Single Clump Studies:  Probing the Fundamental Physics}
\label{sec:single}

The interaction of a radiation flow with a clumpy medium has much broader effects than the alteration of the flow of radiation including outflows and shocks.  The magnitude of these effects depends upon the properties of the clumps.  To understand this physics, we have conducted a number of focused, single clump radiation studies.  For these studies, we use our standard experimentally-motivated target (Section~\ref{sec:simulation}) with a single embedded clump.  These single-clump simulations allow us to study the physical effects and the dependencies of these effects on the properties of the clumps.

\begin{figure}[ht]
\includegraphics[width=7.1in]{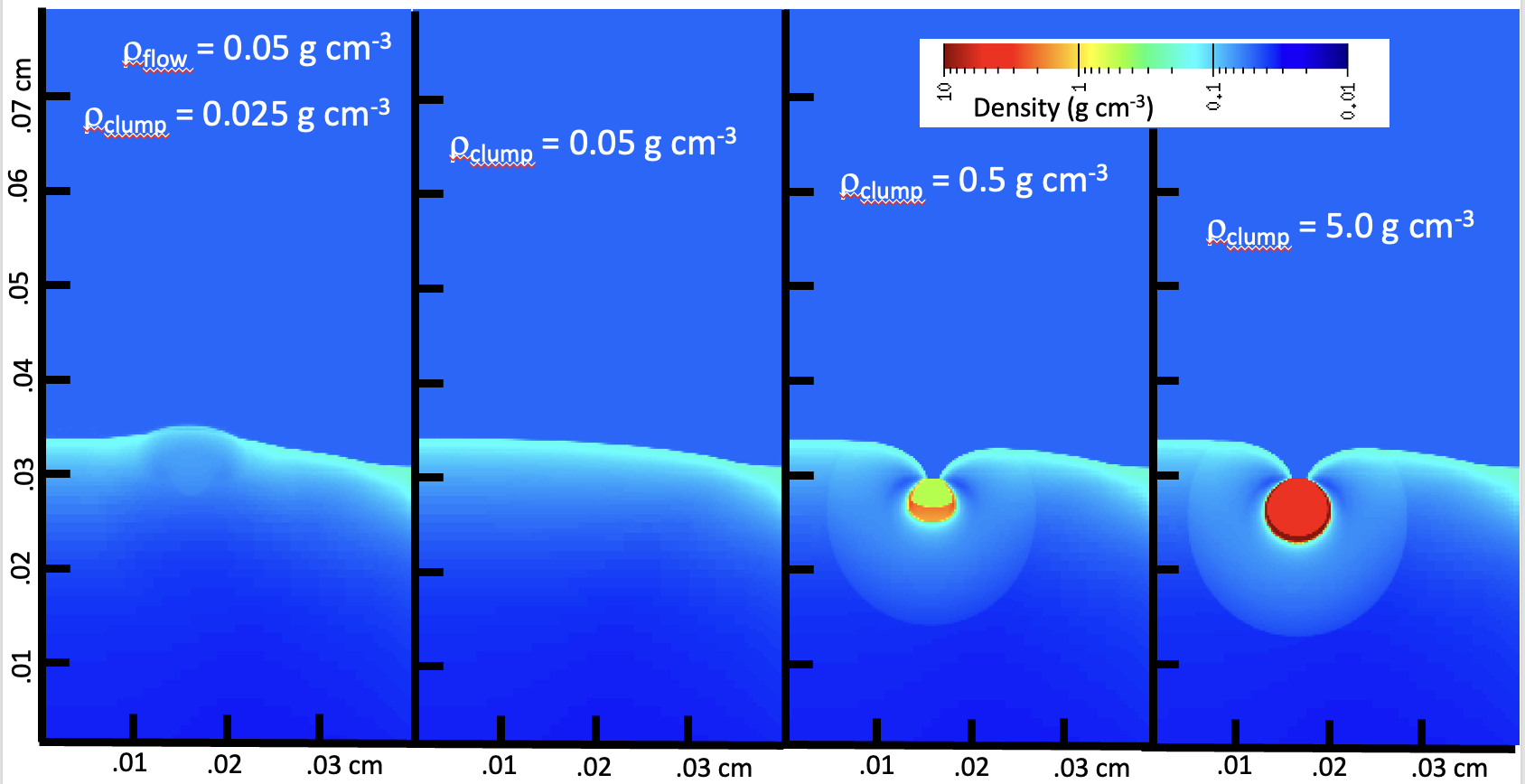}
    \caption{Density map for 4 cylindrical models after 1.6\,ns where the clump density is varied (left to right) from 1/2 of the flow region density to 100 times the flow region density.  The opacity varies dramatically as the radiation front sweeps down the flow region.  For the cold material, the initial Rosseland opacity for the low-energy X-ray photons in this flow is $> 4 \times 10^{4} {\rm \, cm^2 \, g^{-1}}$.  Once heated, the opacity drops by over an order of magnitude and the mean free path in the foam region is roughly 0.01\,cm and the flow region is roughly 8 mean free paths long.  The clump opacity is roughly 50\% higher.  In the high density clump (roughly the density used for most of our calculations), the mean free path is 0.03--0.1~$\mu$m and since the bulk of the clump remains cold throughout the calculation, this mean free path remains low.}
    \label{fig:singleden_den}
\end{figure}

As a first study, we focus on the effects of clump density.  Figure~\ref{fig:singleden_den} shows the density map of 4 separate simulations 1.6\,ns after the launch of the radiation front where the only difference in the 4 simulations is the density of the clump.  For these clumps, the composition of clump and flow regions are identical.  Only the clump density is varied including models where the clump density is lower and higher than the ambient flow region density.  For the low-density clump, radiation rushes through the cavity, leading the flow in the ambient region.  For high density clumps, the radiation flows around the clumps.  For modest density increases (e.g. 10 times the ambient medium), the radiation can both ablate and compress the clump.  However, the timescale to do this increases with increasing clump density and, at the snapshot in time shown in Figure~\ref{fig:singleden_den}, the radiation flow has little effect on the high-density clump.

\begin{figure}[ht]
\includegraphics[width=7.1in]{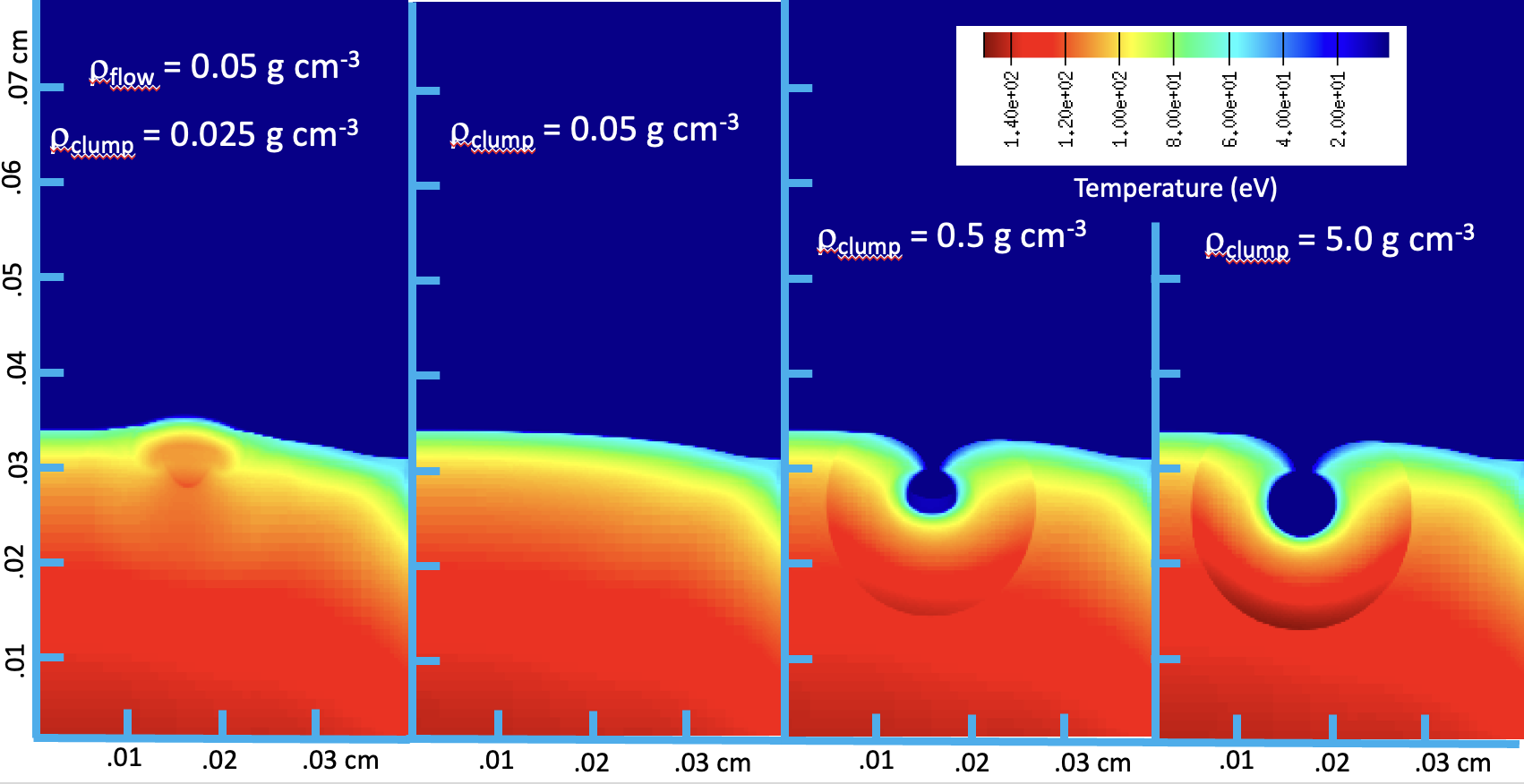}
    \caption{Effective radiation temperature map for 4 cylindrical models after 1.6\,ns where the clump density is varied (left to right) from 1/2 of the flow region density to 100 times the flow region density (same as Figure~\ref{fig:singleden_den}).  Our code models the radiation with 71 energy groups.  The effective radiation temperature is inferred by fitting the spectrum from these groups.}
    \label{fig:singleden_tev}
\end{figure}

The radiation heats up the clump, striving to place it in temperature equilibrium.  If the clump is denser than that of its surroundings, it begins to blow a wind, sending a material flow back into the flow region.  As we shall discuss below, if the composition of the clumps is different than the ambient medium, these outflows mix into the flow region.  If this material has higher opacity, it can choke off the radiation flow. Another feature of these outflows is that they can cause shocks, raising the temperature of the outflow (Figure~\ref{fig:singleden_tev}).  These shocks are critical in understanding supernova outflows, producing extended Ultraviolet and X-ray emission~\citep{2020ApJ...898..123F}.  The alterations in the flow caused by a low-density region can also alter the temperature profile.  Especially in ambient media where the radiation is in the transport or free-streaming regime, this heating from radiation-matter interactions can dramatically alter the emerging spectra.  If the radiation flow is highly photon-energy dependent, this effect can also alter the flow of radiation.

\begin{figure}[ht]
\includegraphics[width=7.1in]{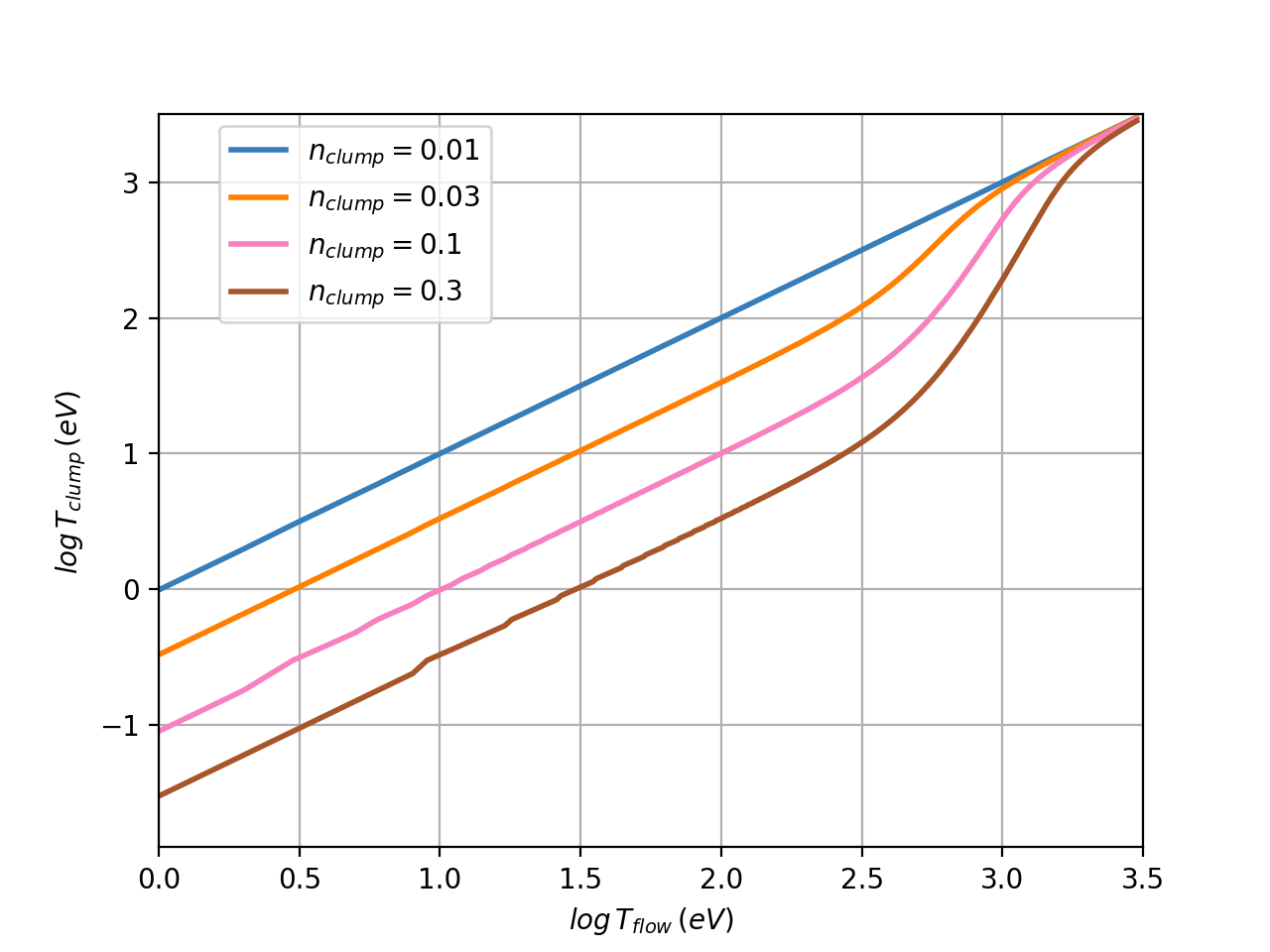}
    \caption{Curves of pressure equilibrium versus the temperature of the flow region (x axis) and the temperature in the clump (y axis).  In this scenario, we assume the number density, $n_{flow}$, for the flow region is $0.01 {\rm \, mol \, cm^{-3}}$ and vary the number density of the clump, $n_{clump}$, over four different values.  These conditions are reasonably close to those in our laboratory experiment.}
    \label{fig:eqpress}
\end{figure}

Matter interactions are important in our simulations.  They drive outflows that both heat the material, altering the energy distribution of the radiation, and inject clump material into the ambient medium.  The physics behind these outflows can be understood through pressure and energetic constraints.  Radiation heats the clumps, striving to achieve temperature equilibrium.  But, as soon as the pressure in the clump exceeds the radiation dominated gas, it will expand.  The pressure in the clump/flow region can be approximated by a combination of ideal gas and radiation pressure:
\begin{equation}
    P_{\rm clump/flow} =  n_{clump/flow} R T_{\rm clump/flow} + \frac{a_{\rm rad}}{3} T^4_{\rm clump/flow}
\end{equation}
where $P_{\rm clump/flow}$, $T_{\rm clump/flow}$ and $n_{\rm clump/flow}$ are the pressure, temperature and number density of the clump or flow region, where the number density is given by the density ($\rho_{\rm clump/flow}$) divided by the average atomic weight of the material ($\langle A \rangle$), $R=8.31 \times 10^7 {\rm erg \, K^{-1} mol^{-1}}$ is the universal gas constant, and $a_{\rm rad}=7.567 \times 10^{-15} {\rm \, erg \, cm^{-3} \, K^{-4}}$ is the radiation constant.  Figure~\ref{fig:eqpress} shows the clump temperature versus the flow-region temperature when the pressure in the clump equals that of the flow region.  The lines in this plot denote pressure equilibrium at the two temperatures on the x and y axes.  We have set up rough conditions for our experiment where the number density, $n_{clump/flow}$,  $= 0.01$ and we vary the clump value up to 30 times higher.  At these densities, ideal gas pressure dominates the radiation pressure at the temperatures in the experiment.  As the clump is heated, its pressure will exceed that of the flow region and it will expand, blowing a wind into the flow region.  

If the radiation pressure dominates (at roughly 1--3\,keV for these conditions), the blow-off is likely to be minimal.  For many astrophysical applications, the density of the material is lower and the temperature at which radiation dominates will also be lower.  The critical temperature ($T_{\rm crit}$) can be estimated by assuming that radiation pressure dominates the flow region and determining the temperature in an ideal gas to match that pressure, i.e. 
\begin{equation}
    T_{\rm crit} = \frac{a_{\rm rad} T^4_{\rm flow}}{3 n_{clump} R} \,.
\end{equation}
Table~\ref{tab:radpres} provides rough values of the critical temperature above which radiation pressure dominates and this blow-off is minimized.  For the different phenomena, both the clump density and temperature of the flow region can vary by many orders of magnitude~\citep{2024A&A...684A.177D}.  For most stars, high stellar radii and low mass-loss rates lead to conditions where the stellar radiation dominates the pressure.  Even though Wolf-Rayet stars are hotter, the mass loss rates are higher and the stars are more compact~\citep{2007ARA&A..45..177C,2020MNRAS.499..873S}.  Many stellar models predict bursts of mass ejection from the star, either through explosive shell burning, opacity-driven instabilities and pressure waves~\citep{2006ApJ...647.1269F,2014ApJ...792L...3H,2016MNRAS.458.1214Q}.  Especially in stripped stars, these can produce prompt ejection rates that are much higher than the mass loss rates of WR stars.  We include all of these scenarios in Table~\ref{tab:radpres}.

\begin{table*}
\begin{center}
\begin{tabular}{|l|c|c|c|c|}
\hline\hline              
Phenomena & $n=\rho/\langle A \rangle$ & $T_{\rm crit}$ & $T_{\rm flow}$ & Dominant Pressure \\
\hline
Stellar Wind &  & & & \\
$ dM/dt = 10^{-7} M_\odot {\rm \,y}^{-1}, \, {\rm radius}=10^{13}\,{\rm cm}$  & $5 \times 10^{-17} {\rm \,mol} \,{\rm cm}^{-3}$ & 200\,K & 10,000\,K & Radiation \\
$ dM/dt = 10^{-6} M_\odot {\rm \,y}^{-1}, \, {\rm radius} = 10^{12}\,{\rm cm}$  & $5 \times 10^{-13} {\rm \,mol} \,{\rm cm}^{-3}$ & 2,000\,K & 10,000\,K & Radiation \\
\hline
Wolf-Rayet Wind &  & & & \\
$ dM/dt = 10^{-5} M_\odot {\rm \,y}^{-1}, \, {\rm radius} = 10^{12}\,{\rm cm}$  & $5 \times 10^{-13} {\rm \,mol} \,{\rm cm}^{-3}$ & 4,500\,K & 20,000--200,000\,K & Radiation \\
$ dM/dt = 10^{-3} M_\odot {\rm \,y}^{-1}, \, {\rm radius} = 10^{12}\,{\rm cm}$  & $5 \times 10^{-11} {\rm \,mol} \,{\rm cm}^{-3}$ & 20,000\,K & 20,000--200,000\,K & Radiation \\
Explosive Mass Ejection &  & & &  \\
$ dM/dt = 10^{-3} M_\odot {\rm \,y}^{-1}, \, {\rm radius} = 10^{11}\,{\rm cm}$  & $5 \times 10^{-9} {\rm \,mol} \, {\rm cm}^{-3}$ & 90,000\,K & 20,000--200,000\,K & Rad/Mat \\
$ dM/dt = 10^{-1} M_\odot {\rm \,y}^{-1}, \, {\rm radius} = 10^{12}\,{\rm cm}$  & $5 \times 10^{-9} {\rm \,mol} \, {\rm cm}^{-3}$ & 90,00\,0K & 20,000--200,000\,K & Rad/Mat \\
\hline
Molecular Clouds & $1.7 \times 10^{-19} {\rm \,mol} \, {\rm cm}^{-3}$ & 30\,K & 10\,K & Mat/Rad \\
\hline
Clouds & $0.056 {\rm \,mol} \, {\rm cm}^{-3}$ & 1.8\,keV & 1\,eV & material \\
\hline\hline
\end{tabular}
\caption{Dominant pressure term for different applications.  Even though radiation dominates many of the applications, we note that blow-off of the clumps may still play a key role in the evolution (see Section~\ref{sec:multi}).}
\label{tab:radpres}
\end{center}
\end{table*}

The specific heat of the clump material dictates the amount of energy that must be injected into this material to heat it ($\Delta T \propto \Delta E/c_v$:  the change in temperature is proportional to the change in energy divided by the specific heat).  This will determine both how quickly the clump begins to expand or blow off as well the temperature of this blow-off. The higher the specific heat, the longer it takes for the radiation and material temperatures to reach an equilibrium.  The specific heat dictates the timescale for the temperature in the clumps to rise.  Because the opacity is extremely sensitive to the temperature, the specific heat can play a huge role in determining the evolution of the opacity in the flow.  

\begin{figure}[ht]
\includegraphics[width=7.1in]{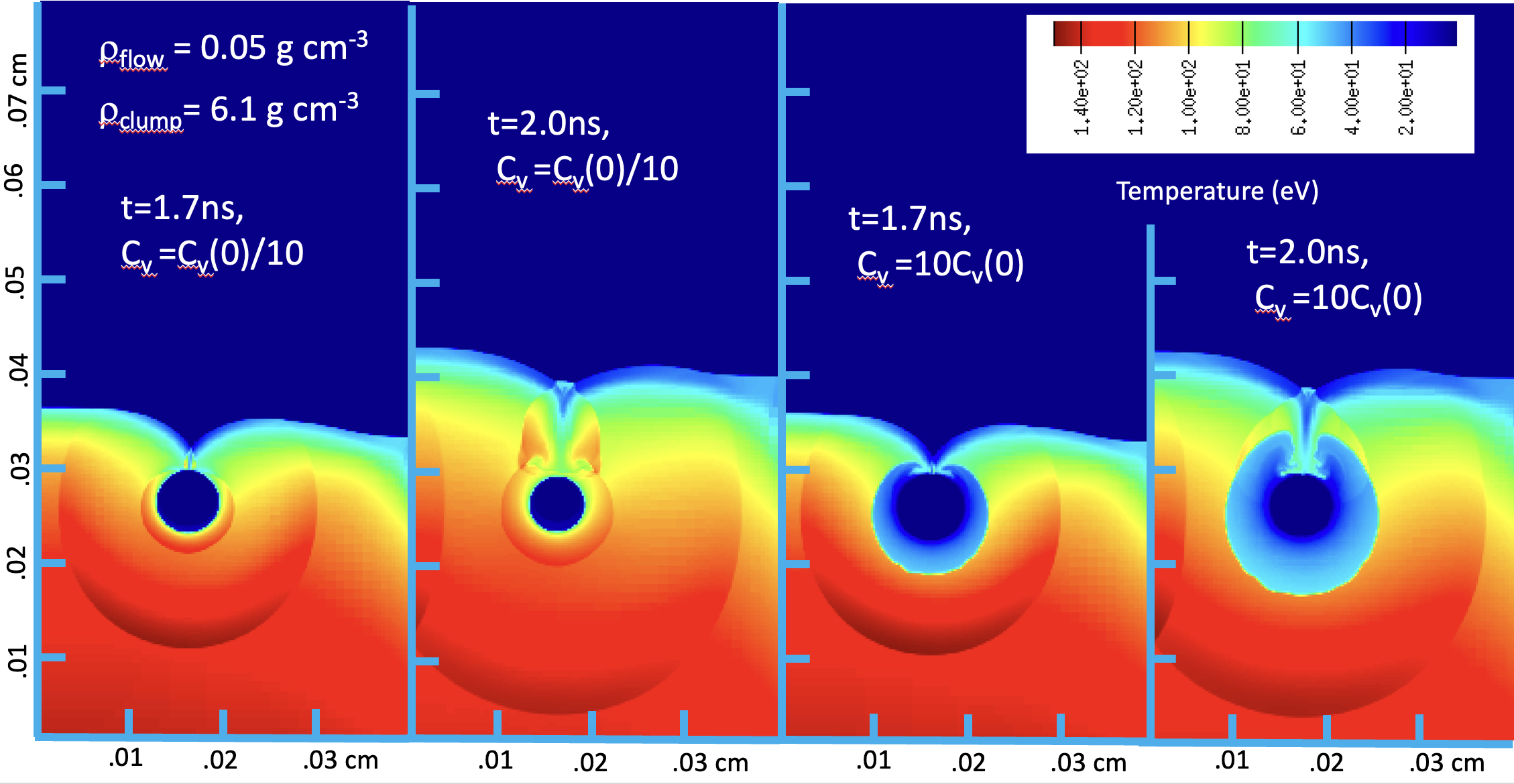}
    \caption{Effective radiation temperature map for two cylindrical models after 1.7 and 2.0\,ns where the specific heat of the clump is lowered and raised by an order of magnitude.  Our code models the radiation with 71 energy groups.  The effective radiation temperature is inferred by fitting the spectrum from these groups.} 
    \label{fig:singlecv_tev}
\end{figure}

From our analytic models, we expect the outflows to depend upon both the specific heat and opacity of the material.  Especially if the the denser clumps are caused through common envelope mass ejection, the clumps are likely to have a different composition than the surrounding wind material.  Although the variation in the specific heat is less important in astrophysical plasmas than in laboratory experiments, opacity differences can be an order of magnitude.  Figure~\ref{fig:singlecv_tev} shows the temperature map from two simulations at two snapshots in time where the specific heat is raised and lowered by an order of magnitude. For high specific heat models, the temperature in the clump need only rise modestly to achieve sufficient pressures to push out against the ambient medium, driving an outflow into the flow region.  If the clump material has a lower specific heat, it must become hotter to achieve the requisite pressures to drive an outflow.  The outflow will take longer to expand and will be hotter.

\begin{figure}[ht]
\includegraphics[width=7.1in]{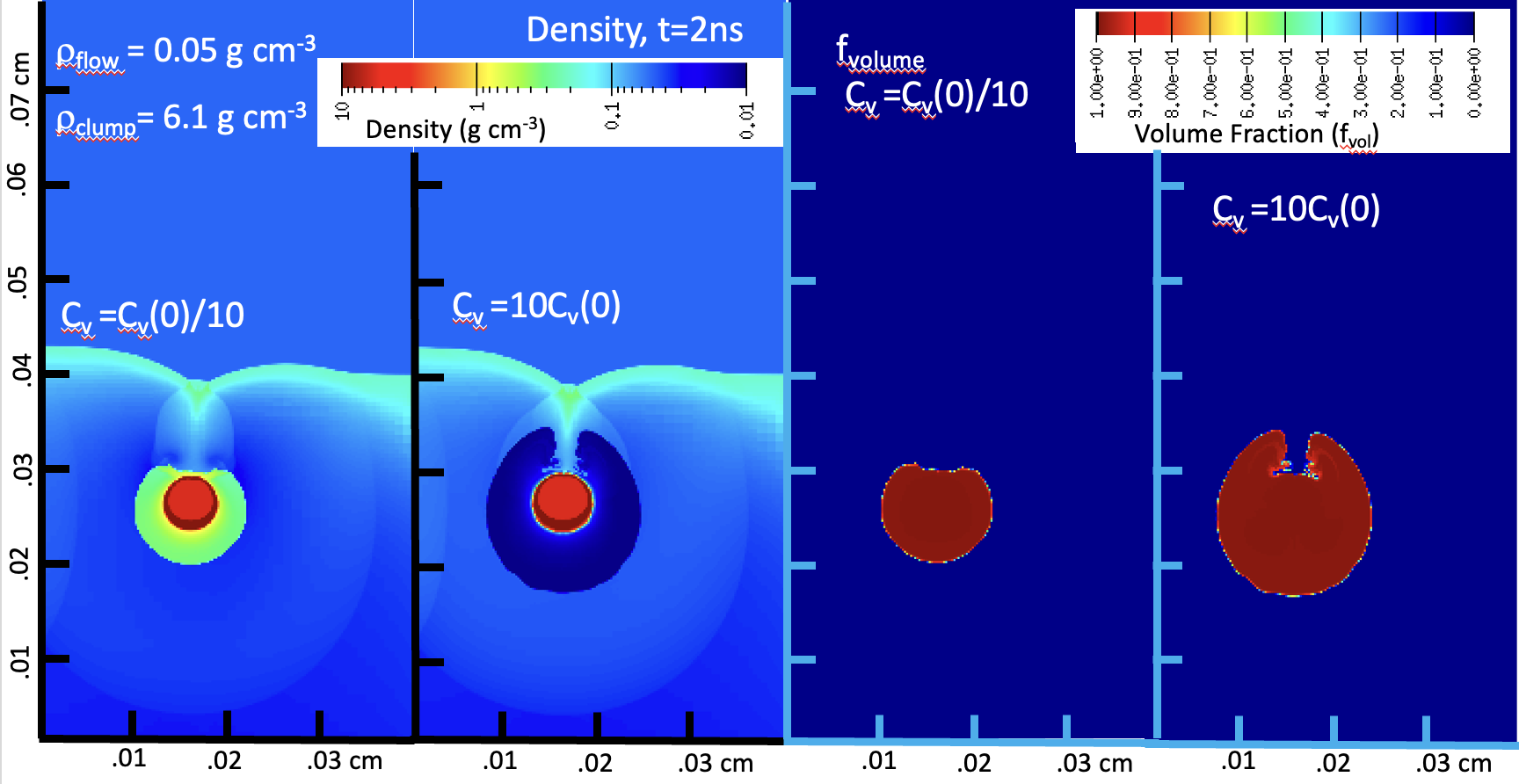}
    \caption{Density and volume fraction ($f_{\rm vol}$) maps for two cylindrical models at 2.0\,ns where the specific heat of the clump is lowered and raised by an order of magnitude.  The density of the outflow from the high specific heat clump material is much lower than that of the low specific heat clump material.  Because the outflow begins earlier for the high specific heat clump material, its expansion is further out.}
    \label{fig:singlecv_denfvol}
\end{figure}

Figure~\ref{fig:singlecv_denfvol} shows the density and volume fraction ($f_{\rm vol}$) maps\footnote{The volume fraction is close to the mass fraction in each zone.  However, {\it Cassio} assumes different materials in a single zone are in Pressure and Temperature equilibrium and the different specific heats of the different materials will lead to different material densities and volume fractions that are slightly different than the mass fractions in a given zone.} corresponding to the temperature maps in Figure~\ref{fig:singlecv_tev}.  Because the outflow starts earlier for the higher specific heat model, it expands more quickly into the ambient medium.  In addition to being lower temperature, the density of the outflow is lower for higher $C_v$ clump.  If the opacity of this material is high, it can cut off the radiation flow.  Although the high specific heat material expands more quickly, it is less dense, so the impact on the radiation flow may be greater for the clumps with lower specific heats.

\begin{figure}[ht]
\includegraphics[width=7.1in]{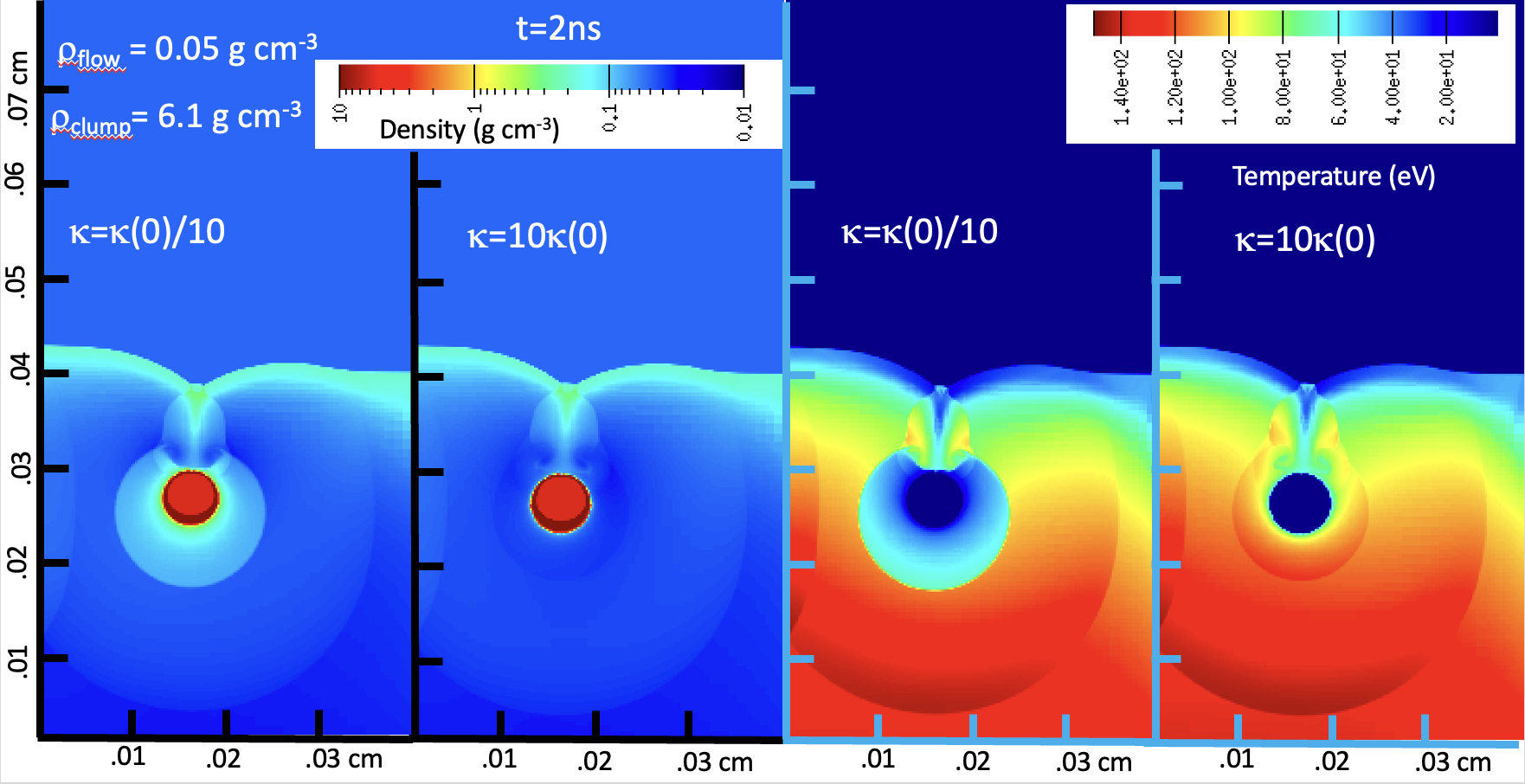}
    \caption{Density and effective radiation temperature maps for two cylindrical models at 2.0\,ns where the opacity of the material in the clump is lowered and raised by an order of magnitude.   As the opacity of the clump increases, the radiation is unable to penetrate the clump leading to a lower-mass wind.  Our code models the radiation with 71 energy groups.  The effective radiation temperature is inferred by fitting the spectrum from these groups. }
    \label{fig:singleop_dentev}
\end{figure}

\begin{figure}[ht]
\includegraphics[width=7.1in]{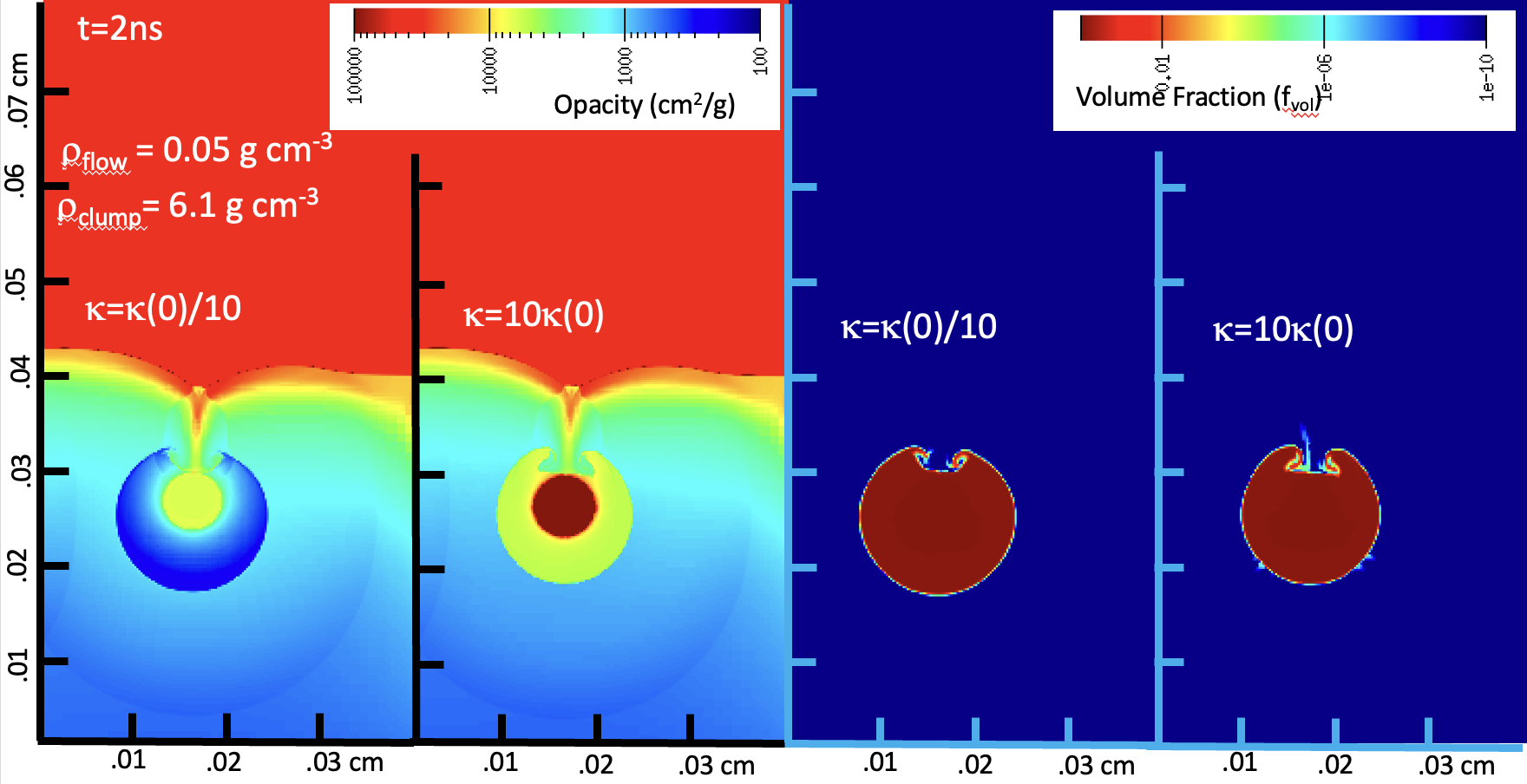}
    \caption{Opacity and volume fraction maps for two cylindrical models at 2.0\,ns where the opacity of the material in the clump is lowered and raised by an order of magnitude.  For the higher-opacity simulation, the mass of the wind is lower, but the opacity of this wind can be much higher.  This will alter the radiation flow.}
    \label{fig:singleop_kapfvol}
\end{figure}

The opacity of the clump material can modify the evolution of the clump.  Figures~\ref{fig:singleop_dentev} and \ref{fig:singleop_kapfvol} show the density, temperature, opacity and volume fraction for models where the opacity of the clump material is raised or lowered.  If the opacity of the clump is high, the radiation is unable to penetrate deeply into the clump.  This means that the wind from the clump taps only a small mass reservoir.  The resulting outflow is lower density but higher temperature.  The extent of the high-opacity outflow is slightly lower.  Depending on how the opacity evolves as it transitions from a solid to a plasma, the nature of the clump-material opacity can have a huge impact on the radiation flow.

One important aspect of many computational studies of transport in inhomogeneous media is that all opacity databases that we are aware of assume that the material is a plasma.  Especially for laboratory experiments, the embedded clumps are in a solid phase, and the opacity of this material can be very different from opacities obtained under plasma conditions.  Electrons obey Fermi-Dirac statistics, which ensures that at solid densities, the partial occupation of higher-lying energy levels decays slowly, resulting in the long characteristic Fermi-Dirac tail. However in dense plasmas, the eigenspectrum is continuous and the partial occupations die out rapidly. This behavior governs the intra (inter) band transitions, or equivalently, bound--bound, bound--free, and free--free electron transitions, and therefore has a direct impact on optical spectra~\citep{Filinov2003, Helled2020}. A recent study of asymmetric carbon--hydrogen (CH) mixtures has shown that compressing the system to three times its density at $k_BT=10$ eV results in an almost threefold increase in its electrical and thermal conductivities~\citep{upcoming}. A compressed ``nonideal" plasma state emerges in numerous dynamic compression environments such as in the laser irradiation of condensed matter in pinched electric discharge experiments, or when the critical density values are attained at high static pressures~\citep{Boehler1993, Aquilanti2015}.  When using laboratory experimental data to test our hydrodynamic feedback to radiative heating, we must understand this physics.

In the presence of spatial inhomogeneities, a subtle interplay exists  between the temperature gradient arising from thermal conduction and the density profile that evolves to compensate for the thermal gradient \cite{Ping2015}.  At an inhomogeneous interface, different opacities create a temperature gradient at the junction driving a density gradient. The response of the density to this induced temperature change can serve as a diagnostic tool for thermal conductivity, a critical parameter for planetary modeling.  

\begin{figure}[ht]
\includegraphics[width=7.1in]{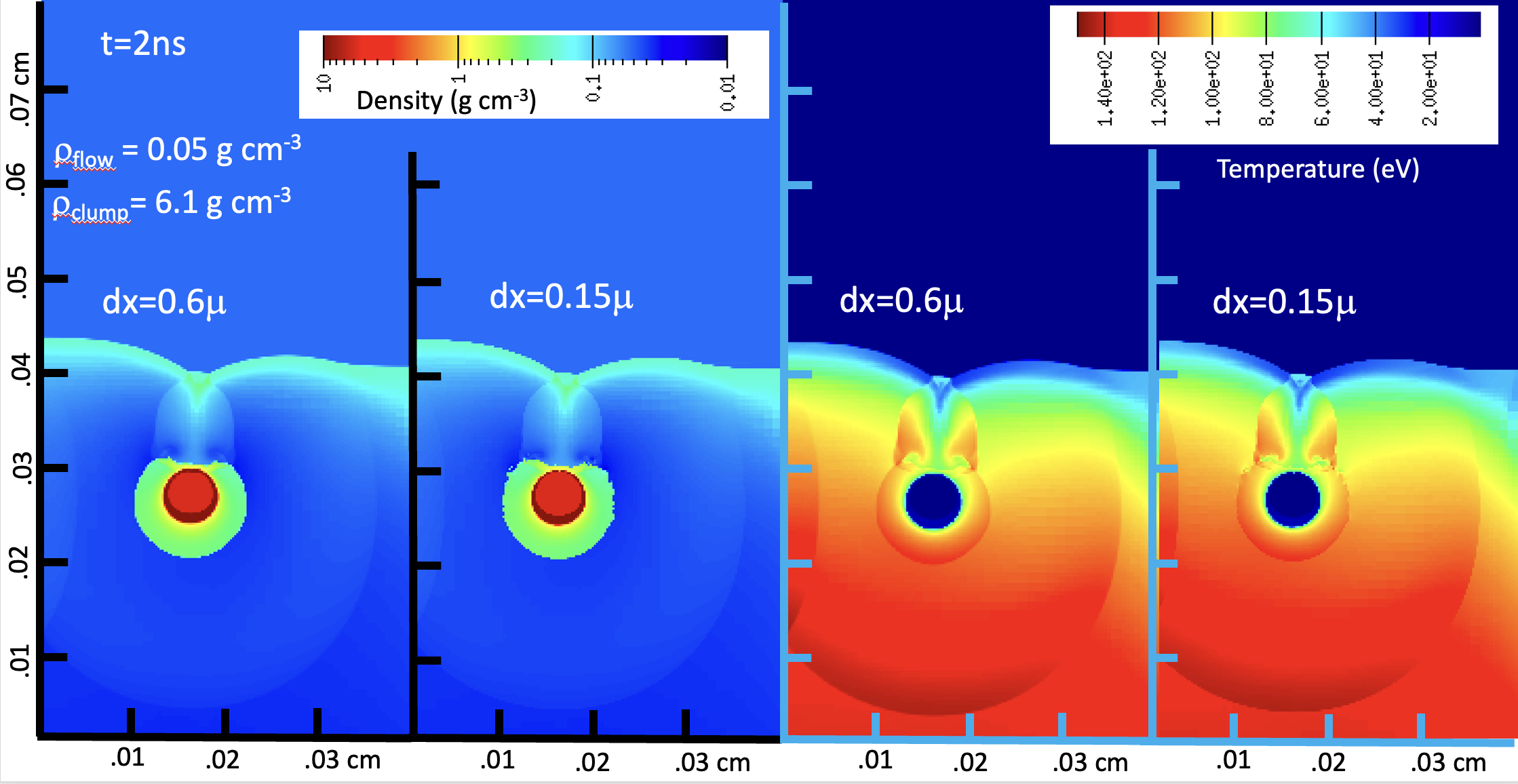}
    \caption{Density and effective temperature maps for two cylindrical models at 2.0\,ns where the resolution is varied by a factor of four (16 times more zones in 2-dimensions). Our code models the radiation with 71 energy groups.  The effective radiation temperature is inferred by fitting the spectrum from these groups.}
    \label{fig:resdentemp}
\end{figure}

In these studies, we have assumed numerical issues (e.g. grid resolution, radiation transport methods) do not significantly alter our results.  Our standard simulations are fairly high resolution (in a $0.04\,{\rm cm} = 400\,\mu m$ target, our resolution is $0.6\,\mu m$).   However, ideally, our calculations would resolve the mean free path of the radiation.  Typical mean free paths in our clump are $0.15\,\mu m$.  To test the resolution in our standard calculations, we ran a series of high-resolution calculations where the resolution was $0.15\,\mu m$.  The results, comparing the standard- to high-resolution calculations is shown in Figure~\ref{fig:resdentemp}.  Although small structures seem to be appearing at $t=2\,$ns, the outflow and shock positions are identical in these calculations. 

\begin{figure}[ht]
\includegraphics[width=3.6in]{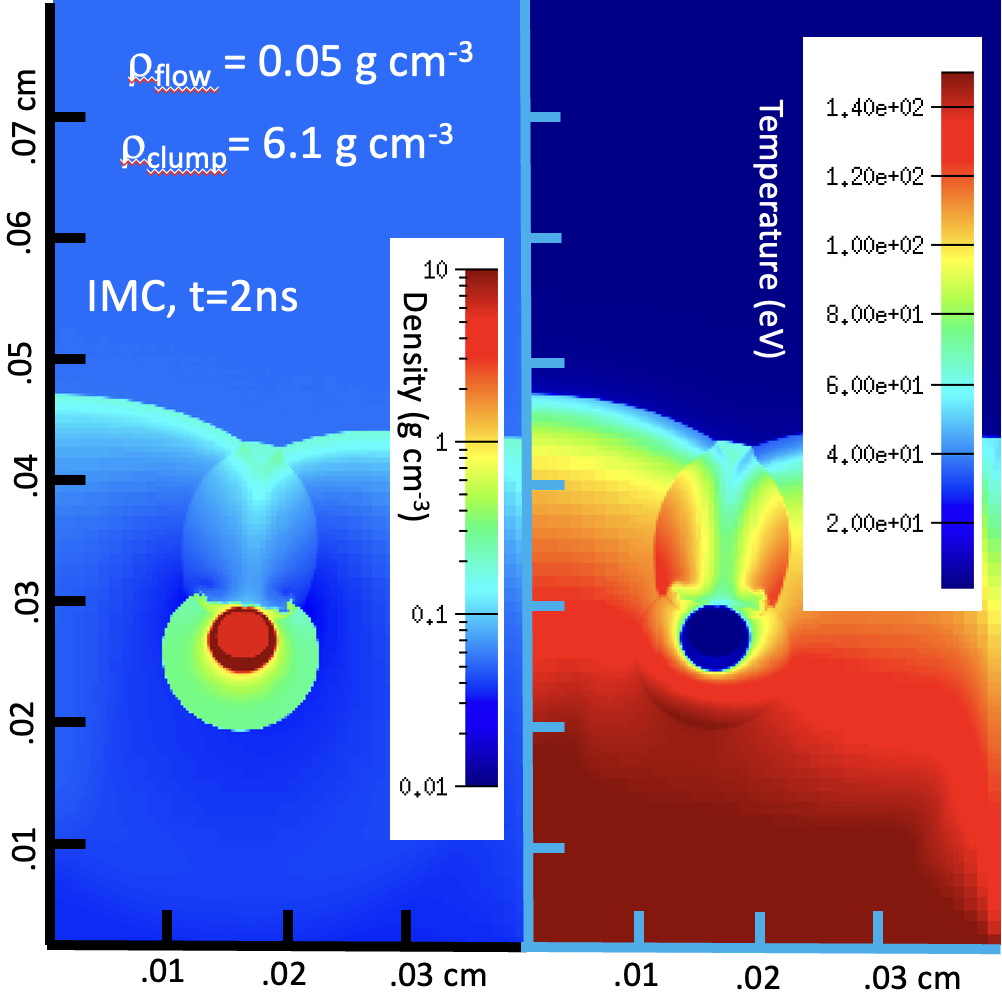}
    \caption{Density and effective radiation temperature maps for our standard cylindrical model at 2.0\,ns using Implicit Monte Carlo. Compare these results to Figure~\ref{fig:resdentemp}.  Despite mild differences in initial conditions and large differences in the numerical approaches, the interaction with the clump with these two schemes is very similar.  Our code models the radiation with 71 energy groups.  The effective radiation temperature is inferred by fitting the spectrum from these groups. }
    \label{fig:imcrun}
\end{figure}

Another numerical uncertainty lies in the implementation of the radiation transport.  As discussed in Section~\ref{sec:simulation}, the {\it Cassio} code includes two higher-order transport methods:  discrete ordinate $S_N$ and IMC methods.  These two approaches use very different mathematical representations of the angular distribution of the photons and the radiation transport.  For most of our calculations, we use the discrete ordinate method.  Figure~\ref{fig:imcrun} shows the results of this IMC run.  The IMC boundary condition is slightly different than that used in the $S_N$ calculations and the angular distribution of the radiation will be slightly different.  Comparing to the results in Figure~\ref{fig:resdentemp}, we note that, although there are differences in the sourcing and in the radiation/material coupling algorithms that deposit energy in the zones, the interaction with the clump and its outflow is very similar between the IMC and $S_N$ calculations. 

One numerical issue that might lead to differences, but that we do not test here, is the implementation of the hydrodynamics.  For instance, the {\it Cassio} code uses an adaptive mesh refinement (we only do 2 levels of refinement).  We do not compare our results to a Lagrangian hydrodynamics scheme.

In some scenarios, the radiation in the flow can cause the clumps to collapse.  For example, in star-forming regions, radiation pressure causes clumps that are on the boundary of gravitational stability to begin to collapse, seeding star formation~\citep{1980SSRv...27..275K}.  Radiation pressure also drives the implosion of capsules in inertial confinement fusion~\citep{1995PhPl....2.3933L}.  The exact details of the collapse depend on the amount of energy injection versus the radiation pressure.  As this subject matter diverges from the primary focus of the studies here, we will defer this discussion and detailed study to a later paper.

Our single-clump observations demonstrate the importance of hydrodynamic feedback in studies of inhomogeneous radiation flow.  In many cases, the clump material can have a different composition than the flow region.  If the clump opacity is very different than the flow region, the blow-off from the heating clump can drastically alter the opacity.  The different line features in the different materials make it possible for laboratory experiments to test the extent of the blow-off.  In supernovae, this blow-off can explain the sudden appearance of lines in the supernova spectra.

\subsection{Multi-Clump Studies:  Putting it all Together}
\label{sec:multi}

Realistic inhomogeneous radiation-flow problems typically involve chaotic inhomogeneities with many dense clumps.  To understand better how radiation flows through a multi-clump medium, we expand our study to include a series of calculations with 25 clumps in our flow region~\citep{byvank23} varied the size and number of clumps in a similar experimental setup).  The simulations shown here all have a size of 0.04\,cm in x-dimension and a y-dimension size of 0.08\,cm.  For our base model, clumps have radii of $50\mu m$.  The mean free path in the flow region ($\lambda_{\rm mfp}^{\rm flow}$) increases as the material heats but is roughly 0.005-0.01\,cm and the mean free path in the clumps is ($\lambda_{\rm mfp}^{\rm clump}$) is $1-2\mu m$.  We also vary the clump size to study the importance of the flow mean free path to the clump size.  To better understand the additional complexity caused by this structure, we have run a series of randomly spaced multi-clump calculations.  We continue to use the clump properties (density, composition) from the XFOL and XFLOWS laboratory experiments~\citep{2023RScI...94b3502J,byvank23}.  We then vary the number, size and distribution of the clumps to better understand the broader implications of inhomogeneous radiation flow.  In particular, these studies allow us to investigate the effects of blow-off (and its dependence on the radiation pressure) under more realistic conditions.  

\begin{figure}[ht]
\includegraphics[width=7.1in]{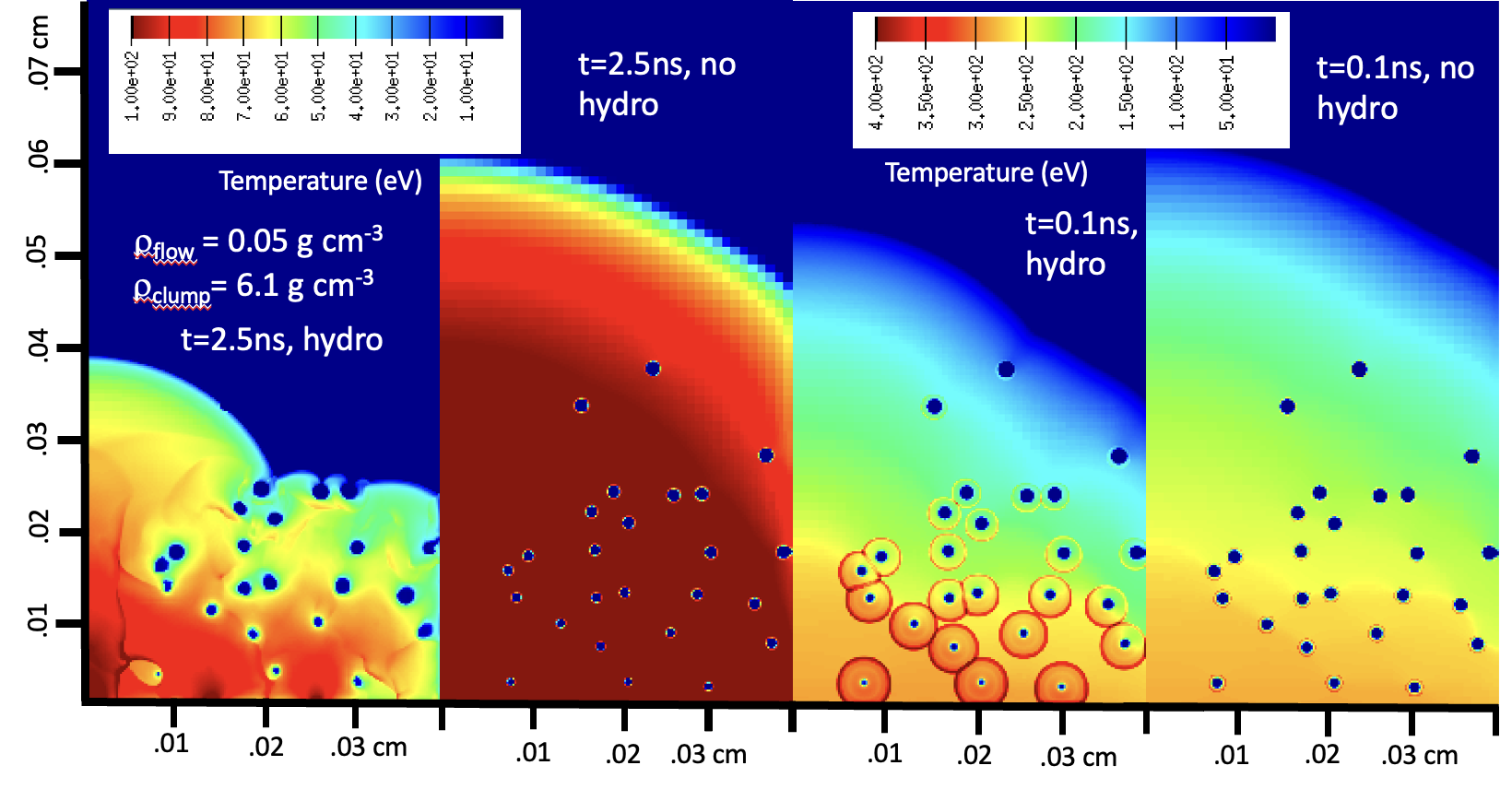}
    \caption{Effective radiation temperature maps for four cylindrical models with 25 material clumps randomly spaced in the flow region.  We use our standard clump and flow densities and compare the solution with and without hydrodynamics turned on.  We also vary the temperature of the drive.  At our standard temperatures, radiative feedback dramatically alters the flow, but, as we increase the temperature, the differences between pure radiation (``no hydro") and radiation-hydrodynamics simulations become smaller.  Our code models the radiation with 71 energy groups.  The effective radiation temperature is inferred by fitting the spectrum from these groups. }
    \label{fig:multitemp}
\end{figure}

For our first study, we focus on the impact of hydrodynamic feedback on the flow of radiation.  Figure~\ref{fig:multitemp} shows the temperature profile of the same initial clump profile, but with two different drives (our standard drive and one where we multiply the temperature of the radiation-driven drive by a factor of 2), each comparing a simulation that is pure transport (``no hydro")) with another that includes hydrodynamic feedback.  The pure-transport calculations show an extreme limit where the radiation simply flows around the clumps.  For the lower-temperature drives, the hydrodynamic feedback is faster than the flow timescale and the clumps expand to block the radiation flow.  The radiation flow is dramatically altered by the blow-off from the clumps.  With a stronger drive, the outflow is minimized due to both higher radiation pressures and the fact that the flow is faster, allowing less time for hydrodynamic feedback.  In the stronger-drive simulations, the differences between the pure-transport and radiation-hydrodynamics simulations are less dramatic.

\begin{figure}[ht]
\includegraphics[width=4.1in]{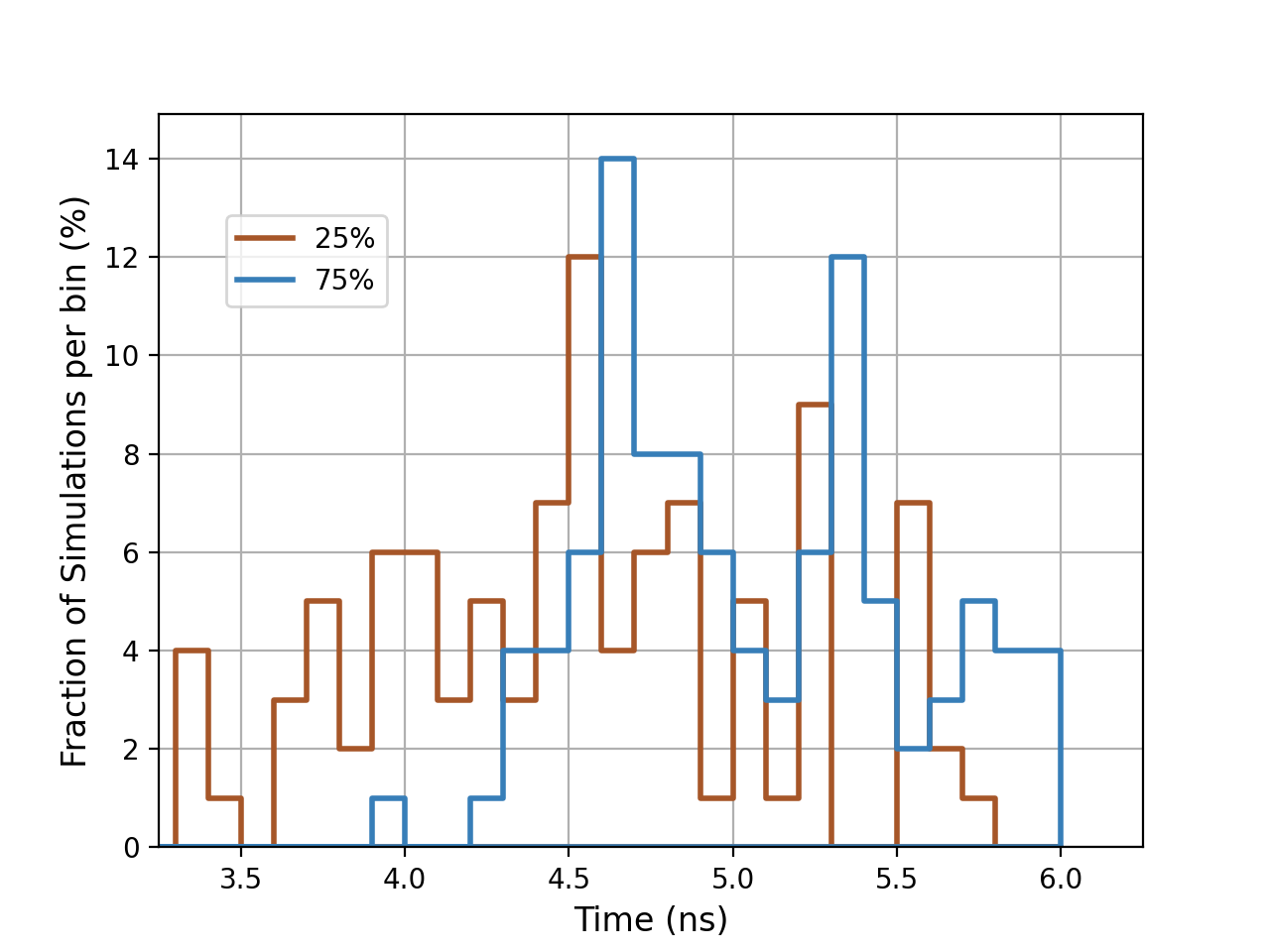}
    \caption{Distribution of radiation-flow timescales for 100 simulations of 25-clump simulations using our standard clump properties ($\rho=6.1 {\rm \, g \, cm^{-3}, 10\,\mu m \, diameter}$), drive power and flow region conditions.  The two distributions correspond to the timescale required for a temperature of 25\% or 75\% of the radiation front at 0.05\,cm above the clumpy medium to rise above 20\,eV.  The spread in the distribution is high (over 2\,ns, $>$ 20\%) and, at least for these 100 simulations, does not appear to be well-matched by a Gaussian.}
    \label{fig:clumpbin}
\end{figure}

Because the radiation flow is strongly affected by the clumps in the material-pressure dominated regime, the flow of radiation strongly depends on the distribution of the clumps (in the low clump-number regime).  We have run 100 different 25-clump calculations, studying the flow of radiation.  Figure~\ref{fig:clumpbin} compares the time it takes for the radiation to emerge from the clump region for 100 distinct simulations, assuming a random distribution of clumps but conserving coverage area.  The propagation timescale of the radiation front varies by 20\%.  For many inhomogeneous flows, it will be difficult to exactly characterize the inhomogeneities and any solution will be limited by this uncertainty.

\begin{figure}[ht]
\includegraphics[width=7.1in]{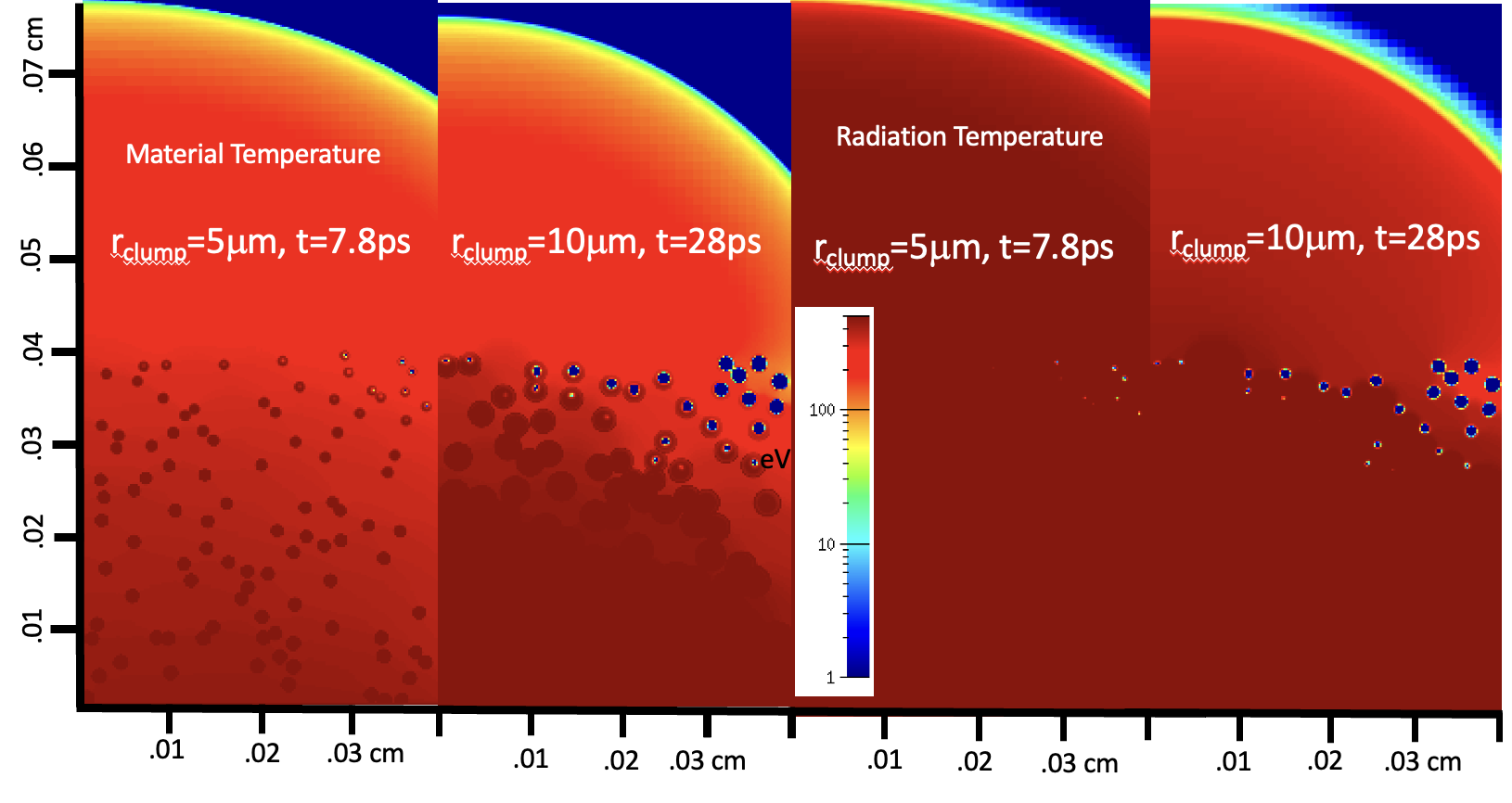}
    \caption{Material and effective radiation temperature for two 100-clump calculations where the radiation temperature drive is 1\,keV.  The two calculations use two different clump sizes ($r_{\rm clump} = 0.05 \mu m, 0.1 \mu m$).  The corresponding clump covering fractions of these two models are 5 and 20\%. We plot the radiation fronts as they reach the edge of our simulation grid, at 7.8\,ps and 28\,ps for the small and large clumps, respectively.  Our code models the radiation with 71 energy groups.  The effective radiation temperature is inferred by fitting the spectrum from these groups. }
    \label{fig:multicltevrev}
\end{figure}

The complex structures are very important in our standard models where the radiation temperature is $\sim 100$\,eV and the material pressure of the clumps (after being heated by the radiation) exceeds the radiation pressure.  We have run a series of simulations using a high-temperature radiation source where the radiation temperature is at 1\,keV.  At these temperatures, radiation pressure dominates, simplifying the radiation flow. However, the physics is still much more complex than simple pressure-equilibrium physics suggests or than the subgrid models discussed in Section~\ref{sec:transportsol} assume.  Figure~\ref{fig:multicltevrev} shows the material and radiation pressure of two 100-clump, high-temperature simulations.  First note that the radiation and material temperatures are very different for these two models.  At these high temperatures, the radiation front moves through our target much faster than the material can equilibrate.  Even without this equilibration, blow-off from the clumps still affects the long-term flow.

\begin{figure}[ht]
\includegraphics[width=7.1in]{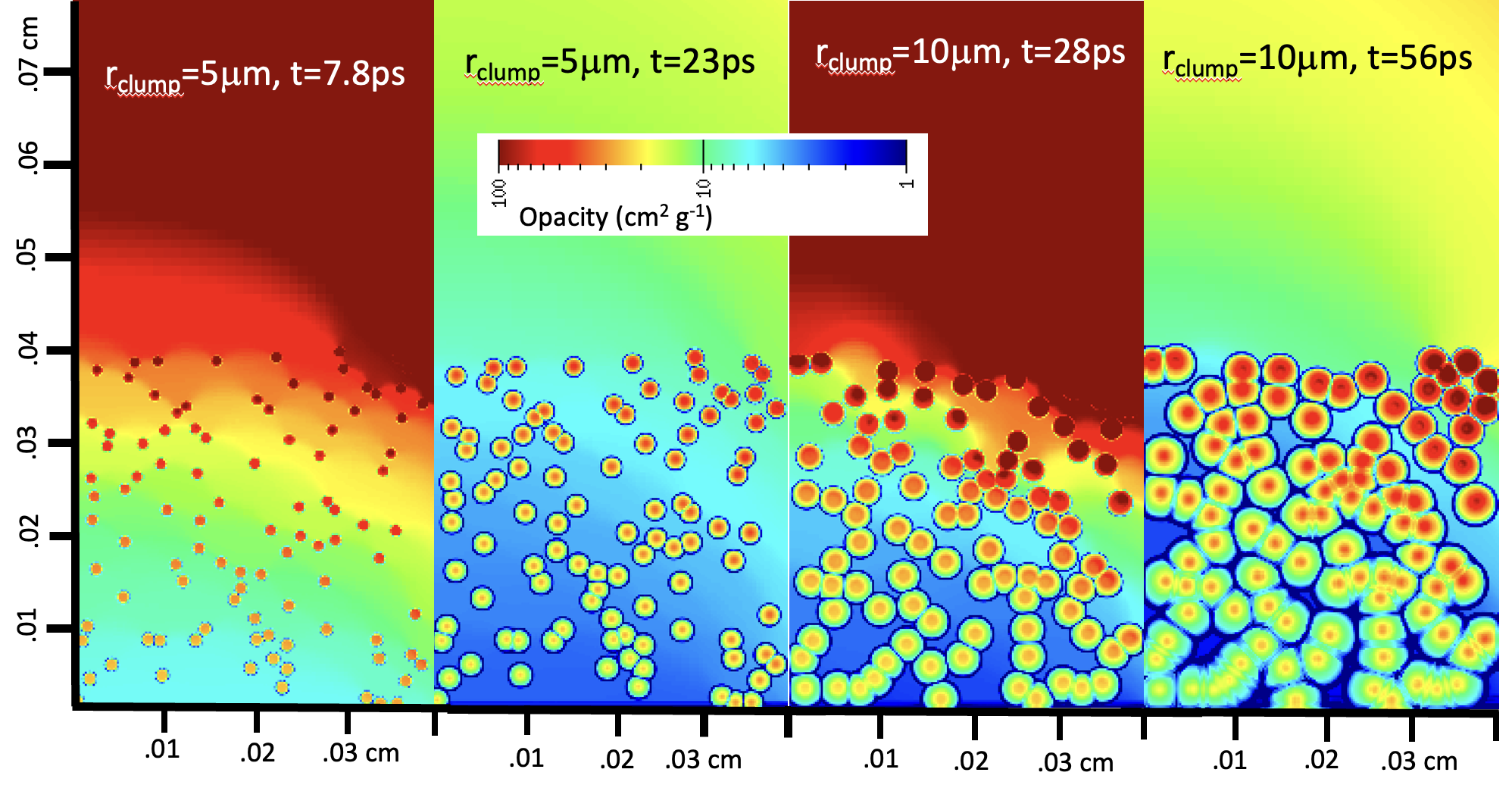}
    \caption{Opacities for the two models from Figure~\ref{fig:multicltevrev} displayed at the two times given in that figure, as well as an intermediate time and a later time, to illustrate the evolution of this opacity with time.  Note that the opacity evolves dramatically with time as the material heats, but the clump opacity evolves much more slowly than that of the flow region.  Also note that, although this is a radiation-dominated flow, the clumps expand and, for the larger clump size, the clump opacities quickly fill most of the region.}
    \label{fig:multicllopa}
\end{figure}

For our clump and flow region compositions, as well as for a fixed density and temperature, we expect the clump opacity to only be a factor of 4 higher than the opacity in the flow region.  As such, following the prescriptions in Section~\ref{sec:transportsol}, we do not expect such a dramatic effect on the propagation timescale.  A few factors contribute to this seeming discrepancy.  First, the material temperatures of the clumps and the flow region are not the same (especially just as the radiation front propagates across the clumps) and this difference becomes larger for larger clumps.  This can lead to a much more dramatic variation between the clump and flow-region opacities (Figure~\ref{fig:multicllopa}).

The opacities for our materials vary dramatically with temperature and this variation is at the heart of the differences between our simulations results and those of the past analytic prescriptions described in Section~\ref{sec:transportsol}.  Because the temperature in the clumps evolves differently than that of the flow region, even if the opacities of the materials are very similar at the same temperature, the opacities for the clump vs. flow region can be very different if their temperatures are different.  Especially near the radiation front, the opacities can differ by more than two orders of magnitude.  In addition, even though the radiation pressure dominates, the clumps do expand and begin to constrain the entire flow region (Figure~\ref{fig:multicllopa}).  If we were only interested in the instantaneous speed of the front, this expansion might not matter.  But in steady-state cases, or cases where the front is driven by further radiation, this clump expansion will alter the radiation flow in this inhomogeneous medium.

\begin{figure}[ht]
\includegraphics[width=5.6in]{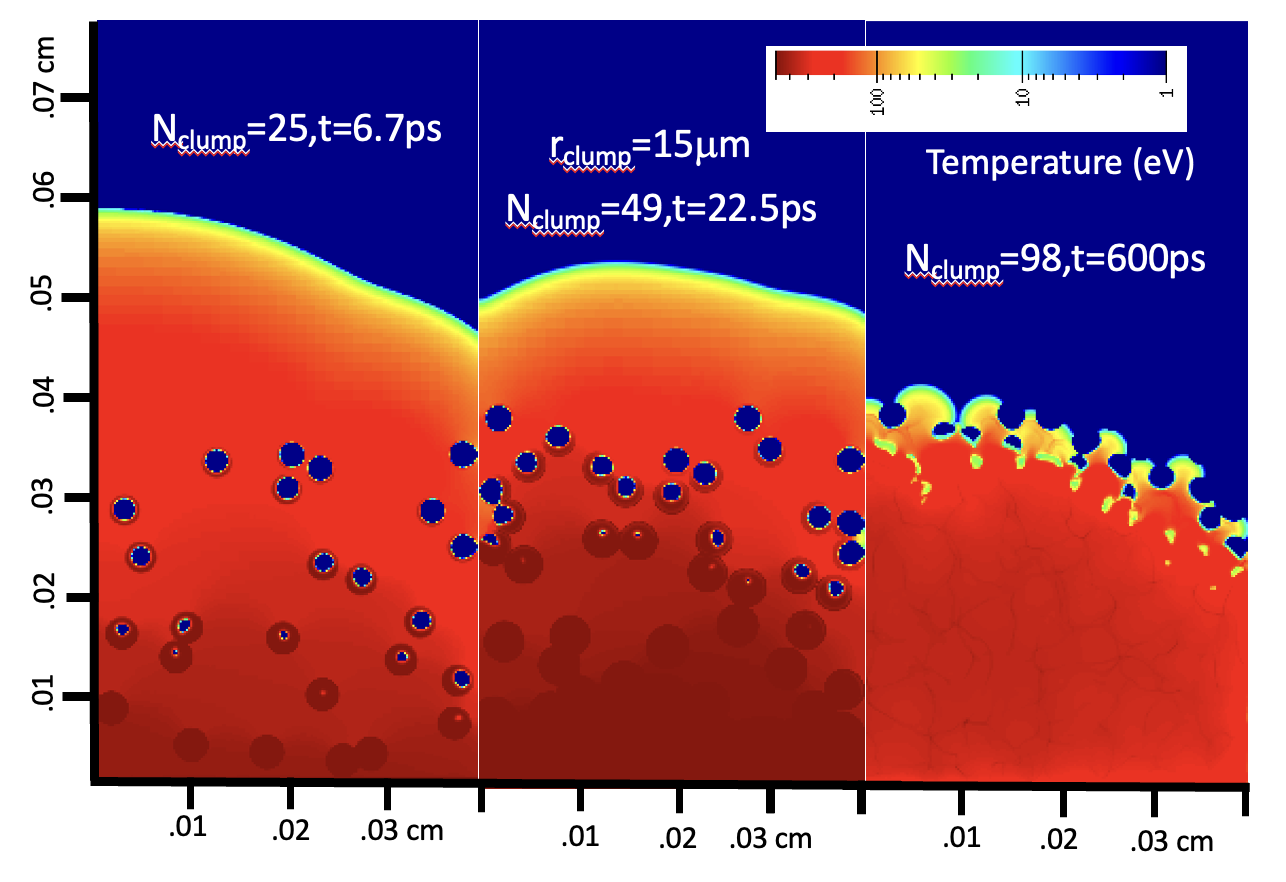}
    \caption{Effective radiation temperature of a strong-drive radiation flow through a clumpy medium where the clump radius is 15\,$\mu$m and the number of clumps varies from 25 to 98.  As the covering fraction increases, the radiation timescale for the propagation of the radiation front increases dramatically and the times for each panel are very different.  Our code models the radiation with 71 energy groups.  The effective radiation temperature is inferred by fitting the spectrum from these groups.} 
    \label{fig:multiclnumcs}
\end{figure}

\begin{figure}[ht]
\includegraphics[width=4.1in]{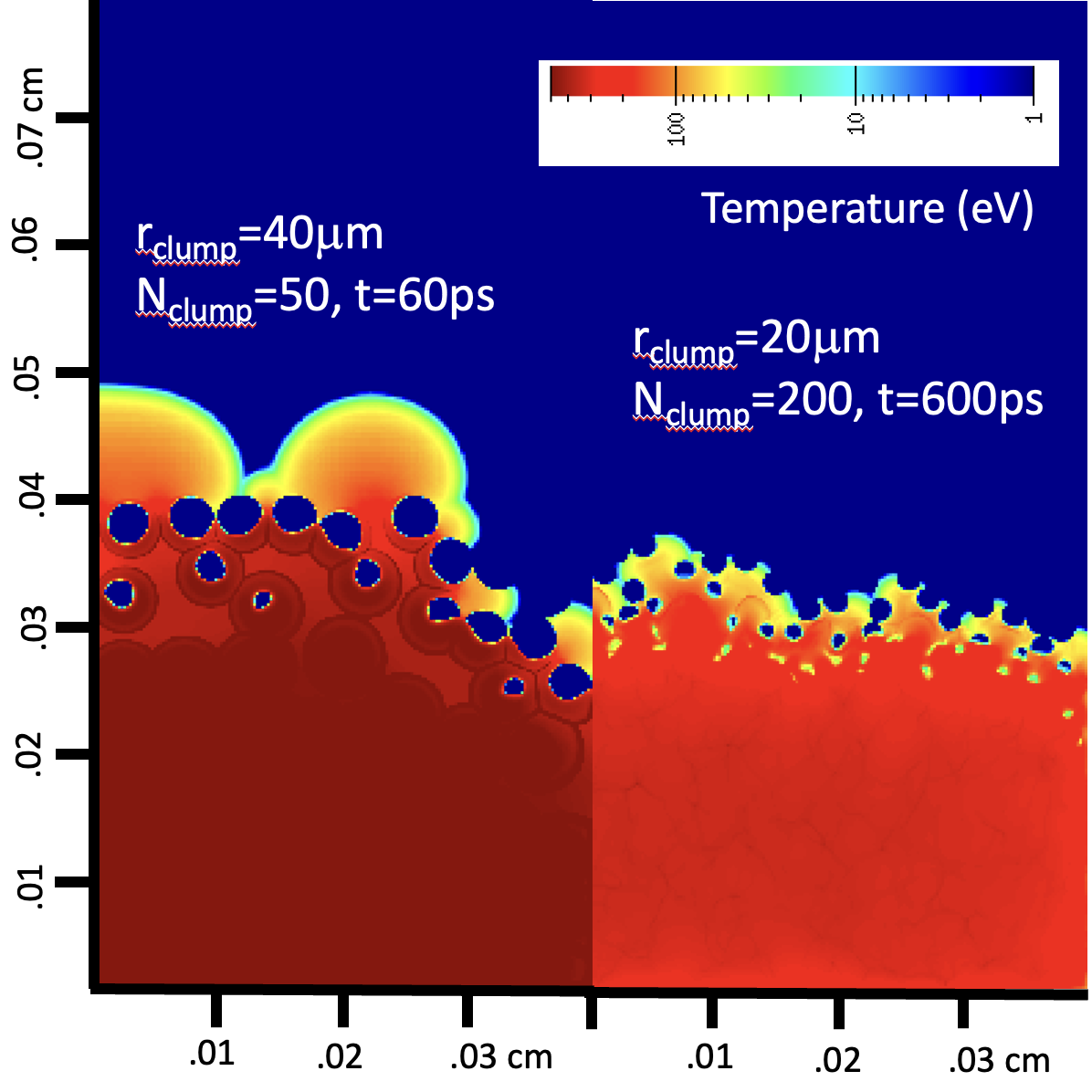}
    \caption{Effective radiation temperature of a strong-drive radiation flow through a clumpy medium where the clump radius and number of clumps are modified to produce the same covering fraction.  Comparing to Figure~\ref{fig:multiclnumcs} demonstrates the importance of the clump size on the radiation flow.  Our code models the radiation with 71 energy groups.  The effective radiation temperature is inferred by fitting the spectrum from these groups. }
    \label{fig:multiclnumsize}
\end{figure}

To further understand the role of clumps, we include two more sets of calculations.  In Figure~\ref{fig:multiclnumcs}, we show the results of three multi-clump calculations, varying the clump number from 25 to nearly 100 clumps.  This set of calculations effectively shows the effect of raising the covering fraction from 11\% to 44\%.  Although the basic trends follow what we expect from our subgrid prescriptions, the opacity evolution and expansion factors exacerbate the effects.  Another way to understand the deviations from our simple subgrid models is to compare the results of two models with the same clump covering fractions, but different size and number of clumps (Figure~\ref{fig:multiclnumsize}).  Although the covering fraction is initially identical between these two runs, the evolution of the radiation fronts are very different.  This behavior is primarily due to the fact that even a small amount of outflow from the clumps can constrain the flow of radiation.  

\begin{figure}[ht]
\includegraphics[width=7.1in]{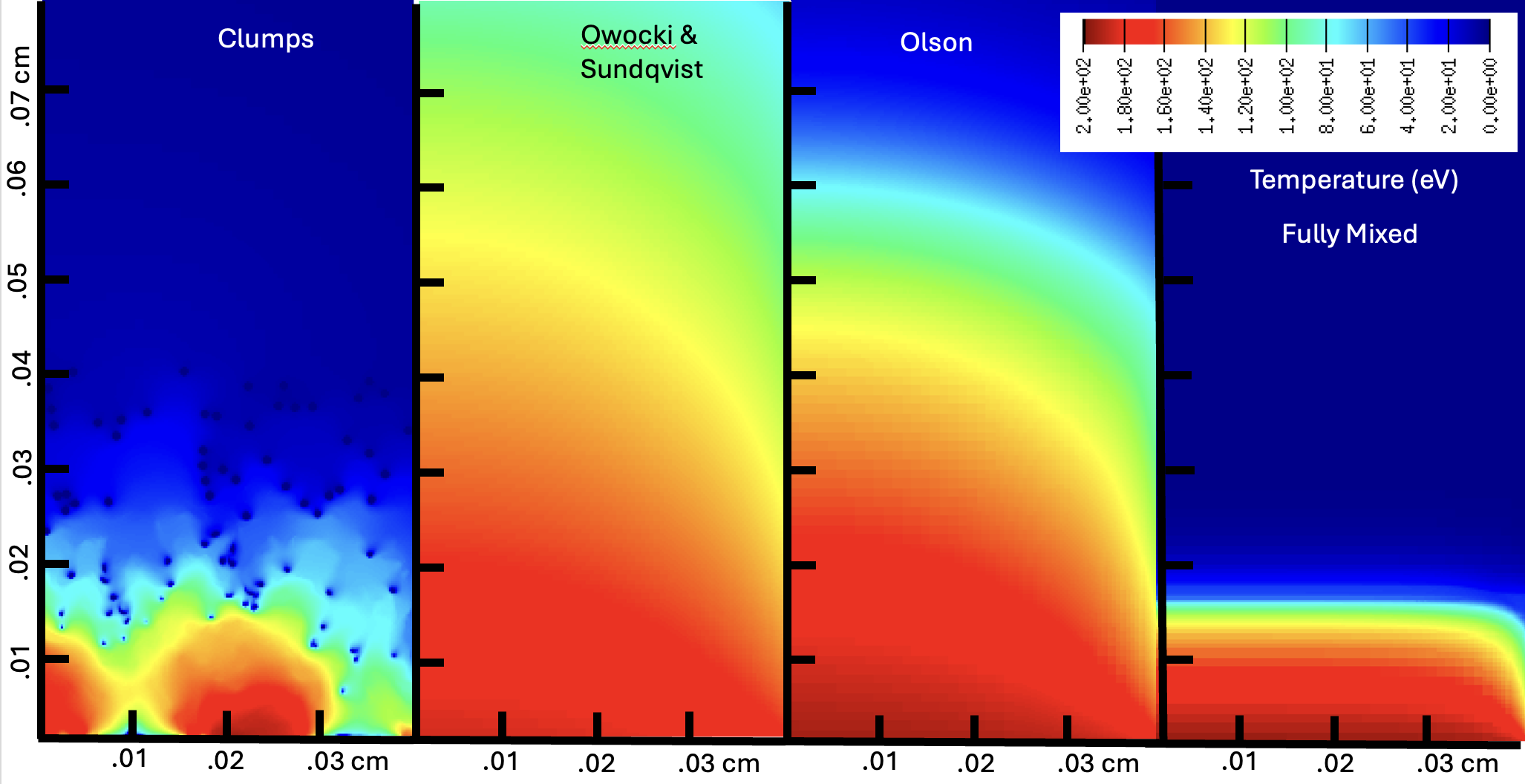}
    \caption{Radiation temperature at 2\,ns for a 100 clump simulation using constant opacities:  $\sigma_{\rm flow}=10^3$, $\sigma_{\rm clump}=10^4$ with corresponding densities of $\rho_{\rm clump} = 6.1\,{\rm g \, cm^{-3}}$and $\rho_{\rm flow} = 6.1\,{\rm g \, cm^{-3}}$.  This simulation is compared to the radiation propagation for both the ``Olson'' (equations~\ref{eq:sigabs},~\ref{eq:sigscat}) formulation and the ``Owocki and Sundqvist'' prescription (equation~\ref{eq:sigowo}). A mass-mixed model (uniformly mixing the clump material throughout the transport regime) is also included in the figure.  This latter solution is a much closer fit to the full solution, likely because the hydrodynamic reaction to the radiation flow causes considerable mixing.}
    \label{fig:opaccomp}
\end{figure}

Clearly, even in the radiation-dominated case, hydrodynamics effects like outflow/winds will be important.  The subgrid formulae in Section~\ref{sec:transportsol} will struggle to match this data.  To test these models, we created a simulation with a simplified opacity prescription, $\sigma_{\rm flow}=10^3$, $\sigma_{\rm clump}=10^4$, within our 100 clump calculation.  The density of the clumps is $6.1\,{\rm g \, cm^{-3}}$ and the flow region is $0.1\,{\rm g \, cm^{-3}}$. The corresponding effective opacity for the ``Olson''~\citep{2005JQSRT..90..131P,2007JQSRT.104...86O} formulation yielded $\sigma_{a,eff} = 1046.2$ (equation~\ref{eq:sigabs}) and $\sigma_{s,eff}=2561$ (equation~\ref{eq:sigscat}.  The corresponding opacity from the ``Owocki and Sundqvist''~\citep{2018MNRAS.475..814O} prescription is $\sigma_{eff}=3949$ (equation~\ref{eq:sigowo}).  Figure~\ref{fig:opaccomp} shows radiation flow calculations at 2\,ns for the clumpy run and these two prescriptions.  For this calculation, a simple mass-mixed model (uniformly mixing, by mass, the clump material in the transport region) is a better fit to the data.

\begin{figure}[ht]
\includegraphics[width=7.1in]{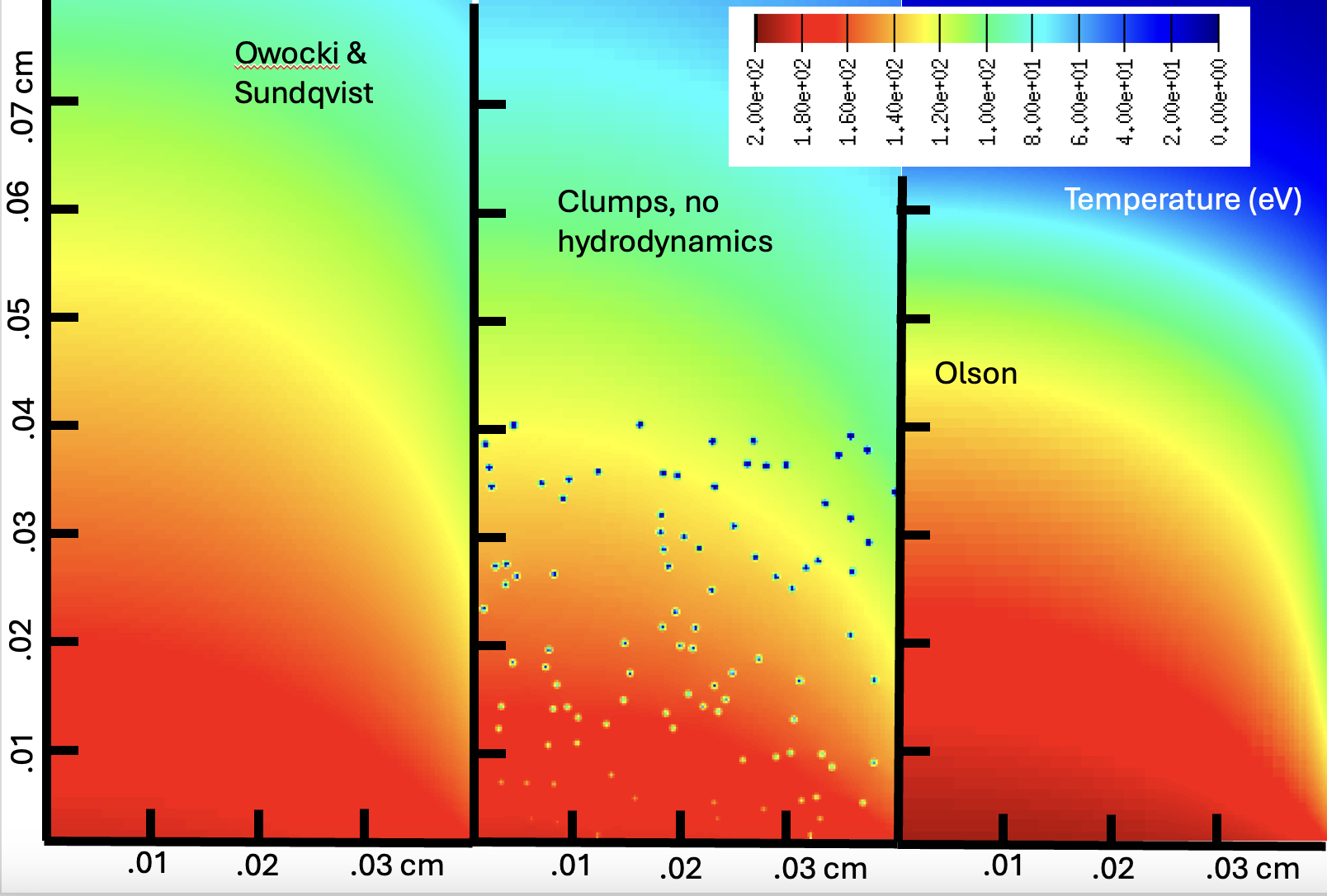}
    \caption{Radiation temperature at 2\,ns for the same simulation as Figure~\ref{fig:opaccomp}.  In this case, the clumpy medium is modeled with the hydrodynamics turned off.  Both of our prescriptions, ``Olson'' and ``Owocki and Sundqvist'' are much better fits to this solution.}
    \label{fig:opaccomp2}
\end{figure}

It is not surprising that the formulations discussed in Section~\ref{sec:transportsol} are not ideal fits to a radiation hydrodynamics solution.  \cite{2005JQSRT..90..131P,2007JQSRT.104...86O} clearly state that they are testing their prescription to a pure transport model.  And, as we have shown, hydrodynamics effects can be extremely important.  If we turn hydrodynamics off, we find in Figure~\ref{fig:opaccomp2} that our prescriptions are a much better fit to the full calculations.  A much more extensive study is required to truly understand the full physics of radiation flow through a multi-clump medium.  We defer this more extensive study, along with a detailed application-specific comparison of these subgrid models to a later paper.

\section{Conclusions}

Radiation flow through inhomogeneous media is critical for a broad range of problems in astrophysics and beyond (see Table~\ref{tab:radpres}).  In many cases, the inhomogeneities can not be resolved in the large-scale calculations of these applications.  In this paper, we reviewed some of the recipes developed to capture radiation transport through these media and the experiments developed to test them.  Existing subgrid models designed to capture unresolved clumps mostly focus on altering the opacity based on the filling factor and opacity of the clumps.  We find that these prescriptions do not match the detailed simulations when hydrodynamic feedback is included.  As we have shown in this paper, the base opacity is also affected by the specific heat of the clumps (because it alters the blow-off) and the size-scale of the clumps.  Any recipe ultimately should include the stochasticity in the final result as well (recall the variation in Figure~\ref{fig:clumpbin}).

This paper included a large set of simulations to better understand the physics behind radiation flow through an inhomogeneous medium.  We found that for regimes where the radiation pressure is not dominant, radiative heating of the high-density clumps can cause the clumps to expand.  This wind can increase the opacity throughout the medium, dramatically altering the flow of radiation.  Most of the recipes in the literature focus on transport-only effects, which are more valid for the conditions where radiation pressure dominates.  But, as we have seen in this paper, even in a radiation-pressure dominated system, radiation-hydrodynamics effects can not be ignored.  In many supernova-interaction problems, this blow-off will play an important role in defining both the broad energy profile and line features of the supernova spectra~\citep{2020ApJ...898..123F,2025arXiv250408889R}.  Radiation will drive interactions prior to the interaction of material shocks and the blow-off material can be more easily swept up in the supernova blastwave.  Several supernova groups are now developing radiation-hydrodynamics codes with higher-order radiation transport schemes to study these effects.  

In all of our models, energy deposition from the radiation front played a much bigger role than momentum deposition and hence, ``outflows" or ``winds" were more important for the subsequent evolution than ablation.  For less dense clumps, as we might expect from turbulent instabilities, ablation can play a much more important role~\citep{2020ApJ...898..123F}.  Ablation will have a similar effect to blow-off in the sense that it will disperse the clumpy material, causing shocks and cutting off the flow.  Ablation may be more important in stellar winds and protostellar outflows.  But the nature of ablation will be different than the simulations shown here and much more work must be done to understand all aspects of inhomogeneous radiation flow for the broad set of applications discussed in this paper.  

In most cases, the hydrodynamic feedback effects are not fully captured by current prescriptions designed to capture a modified total opacity or one that incorporates an effective scattering term to capture angular effects.  Especially in problems studying detailed spectra (e.g. in laboratory experiments or supernova light-curves), the blow-off mixing and shock interactions must also be captured by any subgrid prescription.

The experiments can be developed to produce scalable rations of the mean free path ($\lambda_{\rm mfp}^{\rm flow}$), clump mean free path ($\lambda_{\rm mfp}^{\rm clump}$) and clump size ($d^{\rm clump}$):  e.g.  $\lambda_{\rm mfp}^{\rm flow}/ \lambda_{\rm mfp}^{\rm clump}$, $\lambda_{\rm mfp}^{\rm flow}\rm/d^{\rm clump}$.  These allow us to probe the radiation flow in transport conditions that mimic those of a broad range of applications.  In addition, these experiments can study our numerical models to couple radiation and hydrodynamics to determine the uncertainties due to numerical effects like numerical diffusion (both in matter and radiation).

\begin{figure}[ht]
\includegraphics[width=4.1in]{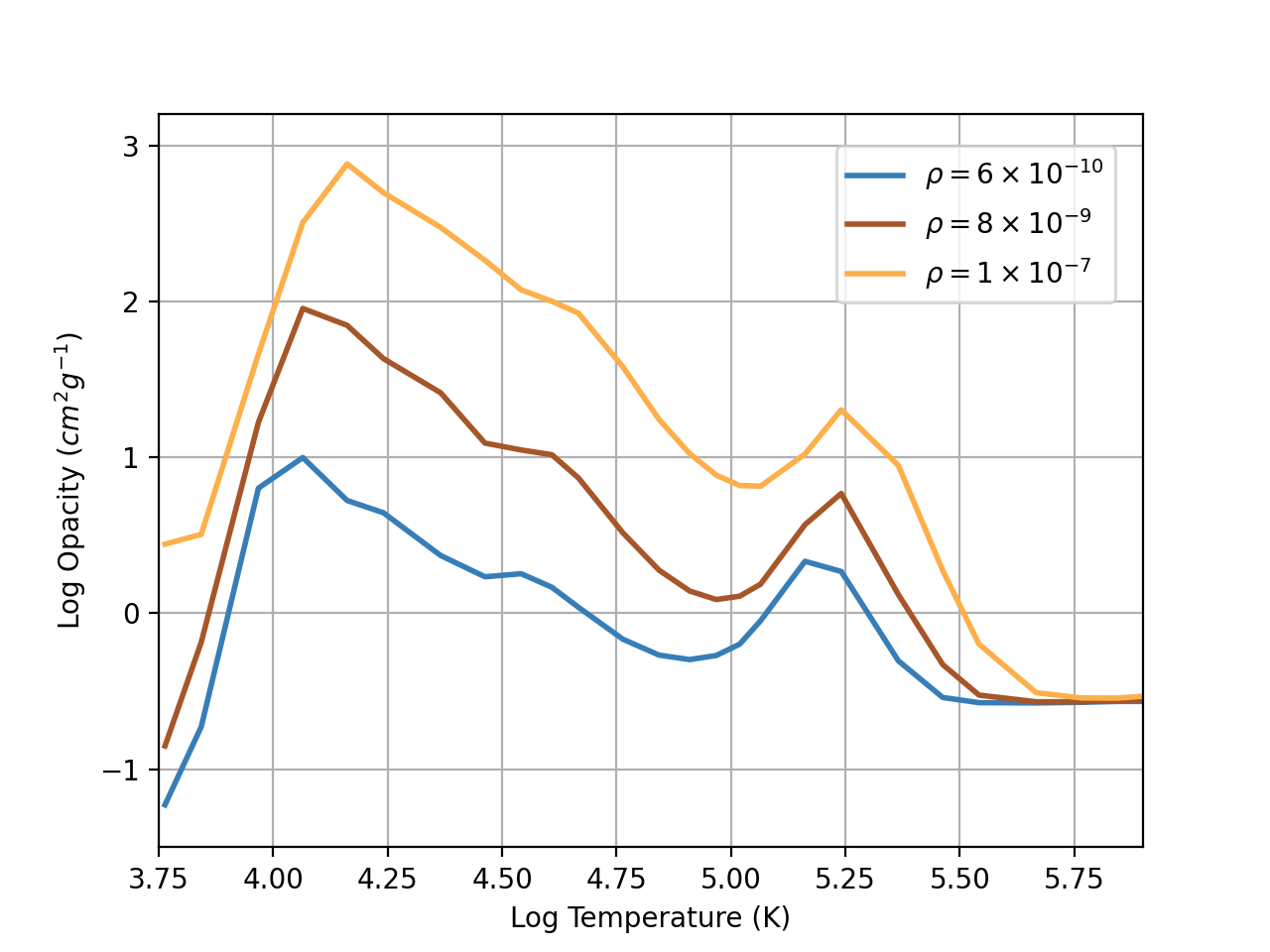}
    \caption{OPLIB opacities versus temperature for solar-abundance material near the edge of a massive star corresponding to the conditions during the onset of shock breakout.  The opacity can drop over three orders of magnitude as the temperature rises from 10,000\,K to a few 100,000\,K.}
    \label{fig:opacsbo}
\end{figure}

We noted the complications caused by opacities that vary rapidly with temperature.  For our materials, the opacity for both our materials drops by over an order of magnitude as the radiation front propagates across the flow region.   The opacity of the clump material will drop much slower because it takes time for it to heat.  The same is true for many astrophysical conditions.  Figure~\ref{fig:opacsbo} shows the solar opacities from the Los Alamos OPLIB database for a range of temperatures and densities.  The temperature of a giant star is roughly $10,000$\,K, but can rise up to 1 million K when the supernova blastwave breaks out of the star.  In such scenarios, that opacity can drop as much as 3 orders of magnitude.  Any subgrid model estimating the opacities of a clumpy medium must include this temperature dependence.  This figure assumes the radiation and matter are in equilibrium.  As we showed in our simulations, it is unlikely that this will hold in many applications.  Out-of-equilibrium effects must be included to fully follow this radiation flow.

At this time, the physics is sufficiently complex that no model captures all of these effects.  A much more comprehensive study is needed to develop a more generic solution and it may be that the more appropriate approach would be to devise individual recipes for specific problems.  We defer this study to a later paper.

\begin{acknowledgements}

This work was supported by the US Department of Energy through the Los Alamos National Laboratory. Los Alamos National Laboratory is operated by Triad National Security, LLC, for the National Nuclear Security Administration of U.S.\ Department of Energy (Contract No.\ 89233218CNA000001).  A portion of the work by CLF was performed at the Aspen Center for Physics, which is supported by National Science Foundation grant PHY-1607611.

\end{acknowledgements}

\bibliography{refs}{}
\bibliographystyle{aasjournal}

\end{document}